\documentclass[11pt,a4paper]{article}
\usepackage{graphicx} % Required for inserting images
\usepackage{amsmath}
\usepackage{jheppub}
\usepackage{subcaption}
\newcommand{\figref}[1]{Fig.~\ref{#1}}
\newcommand{\secref}[1]{Section~\ref{#1}}
\newcommand{\appref}[1]{Appendix~\ref{#1}}
\newcommand{\eqnref}[1]{Eq.~\ref{#1}}

\newcommand{\micromegas}{\texttt{micrOMEGAs}}
\newcommand{\madgraph}{\texttt{MadGraph}}
\newcommand{\feynrules}{\texttt{FeynRules}}

\title{Light Vector Dark Matter via a Magnetic Dipole Portal: Bridging Direct Detection and Fixed-Target Searches}
\author[a]{Avik Banerjee,}
\author[b]{Riccardo Catena,}
\author[b]{and Taylor R. Gray}

\affiliation[a]{Department of Theoretical Physics, Tata Institute of Fundamental Research, Homi Bhabha Road, Mumbai 400005, India}
\affiliation[b]{Chalmers University of Technology, Department of Physics and Astronomy, SE-412 96 G\"oteborg, Sweden}

\emailAdd{avik.banerjee\_205@tifr.res.in}
\emailAdd{catena@chalmers.se}
\emailAdd{taylor.gray@chalmers.se}

\abstract{
We present a model featuring sub-GeV vector dark matter by augmenting the Standard Model with a new non-Abelian dark $SU(2)_D$, spontaneously broken by the vacuum expectation values of a scalar doublet and a triplet. Interactions between the dark and visible sectors arise through a dimension-5 non-Abelian \textit{avatar} of kinetic mixing portal, inducing effective magnetic dipole couplings of the dark matter, with the photon and Z boson. The resulting spectrum of the dark gauge bosons naturally exhibits an inverse mass hierarchy between the dark matter and the $Z^\prime$, leading to interesting phenomenology at fixed-target experiments such as LDMX through dark off-shell bremsstrahlung, dark Higgs-strahlung, invisible vector meson decay, and visible decays. We compute the thermal relic abundance across sub-GeV dark matter masses, with regions of freeze-out proceeding via annihilation into dark sector states or direct annihilation into Standard Model states. Bounds from a broad set of laboratory probes, along with cosmological and astrophysical observations, are incorporated in our analysis. Among them, the most restrictive bounds originate from direct detection experiments, Big Bang Nucleosynthesis, collider searches, and the CMB. Our results demonstrate that a sizeable region of parameter space remains consistent with the observed relic abundance and current experimental constraints, and that fixed-target searches should be considered in tandem with direct detection and cosmological probes for an effective and comprehensive search strategy, especially in the off-shell regime.
}

\begin{document}

\maketitle

\section{Introduction}

The quest to uncover the elusive nature of dark matter (DM), strongly supported by evidence across a wide range of cosmological and astrophysical observations \cite{Balazs:2024uyj}, continues along several complementary experimental and theoretical directions. Major global efforts are underway to: (i) perform \textit{direct detection} searches aiming to observe interactions of the Galactic DM at terrestrial detectors; (ii) infer DM properties through its imprint on \textit{cosmological} observables such as the cosmic microwave background (CMB); (iii) identify possible \textit{astrophysical} signals of DM annihilation or decay into Standard Model (SM) particles via \textit{indirect detection} probes; (iv) constrain DM interactions and their effects on the SM through \textit{high-precision measurements} of the SM parameters; and (v) explore the production of DM and dark sector states in \textit{collider} experiments.

Along many of these fronts, the weakly interacting massive particle (WIMP) \cite{Arcadi:2017kky,Steigman:1984ac} has long served as the leading DM candidate, primarily due to its ability to naturally reproduce the observed relic abundance through weak-scale interactions. Although the WIMP paradigm remains appealing, growing attention has turned toward alternative possibilities. In particular, a broad class of models now considers DM candidates with masses below the GeV scale. Such light DM scenarios require the presence of a new mediator particle to facilitate interactions between the DM and the SM, thereby avoiding the Lee--Weinberg bound---an upper mass limit for thermal relics with weak-scale couplings \cite{Lee:1977ua}. 

Since sub-GeV DM is too light to produce detectable nuclear recoils in conventional direct detection experiments, novel search strategies have been developed. A prominent approach involves direct detection of DM--electron scattering in target materials \cite{Mitridate:2022tnv,DAMIC-M:2025luv}, which provides optimal sensitivity to light DM. Concurrently, accelerator-based experiments offer complementary avenues to probe light DM, either through missing energy or momentum signatures, direct DM scattering in downstream detectors, or via visible decays of the mediator into SM particles \cite{Izaguirre:2015yja,Krnjaic:2022ozp}. Several current and upcoming experiments are specifically designed to explore this light DM regime \cite{Balan:2024cmq}.

Fixed-target experiments are particularly well suited for probing light DM owing to their well-characterized backgrounds, high luminosities, and favorable forward kinematics \cite{Berlin:2020uwy}. The forthcoming \textit{Light Dark Matter eXperiment (LDMX)}---a missing-momentum experiment with a planned 8\,GeV electron beam incident on a tungsten target---is projected to substantially extend the current sensitivity reach \cite{LDMX:2025pkp}. Theoretical studies relevant to LDMX and similar light DM searches have predominantly focused on simplified models featuring a dark photon mediator ($A'$) \cite{Berlin:2018bsc} as well as spin-1 DM scenarios encompassing both simplified and extended frameworks \cite{Catena:2023use}. More recently, dark photon models incorporating higher-order dark electromagnetic moments have also been explored \cite{Catena:2025fsl}. 

In general, any heavy new particle that can be produced in electron--nuclear collisions, such as through dark bremsstrahlung, and subsequently escape detection due to its feeble interactions with the SM, constitutes a viable target for LDMX. This broad experimental reach renders LDMX a promising platform to explore a diverse landscape of theoretical possibilities within the dark sector. Among the many possibilities, spin-1 (vector) DM \cite{Hambye:2008bq,Choi:2019zeb,Hu:2021pln,Nomura:2020zlm,Nomura:2021aep,Nomura:2021tmi,Ramos:2021txu,Abe:2020mph,Chu:2023zbo} has been explored in diverse contexts, ranging from direct detection phenomenology \cite{Krnjaic:2023nxe,Catena:2019hzw,Catena:2018uae} to gravitational-wave signatures \cite{Benincasa:2025tdr,Belyaev:2025zse} and freeze-in production mechanisms \cite{Krnjaic:2022wor}. Previous studies, such as Ref.~\cite{Catena:2023use}, have focused on the on-shell production of spin-1 DM at fixed-target facilities like LDMX.  

The commonly studied scenario is the one in which a dark photon is produced on-shell through dark bremsstrahlung, and later decays into DM which goes undetected in the calorimeters \cite{Berlin:2018bsc}. With on-shell $A'$ production, the physics of the missing-momentum signature is independent of the specifics of the DM theory -- only the dark photon properties (such as its kinetic mixing and mass) can alter the signature. If the DM is produced off-shell, however, not only will the signature be suppressed but it will also be dependent on the specific DM theory, since the Feynman diagram contains the outgoing DM fields and associated vertex.

The present work examines a spin-1 DM scenario featuring  a complementary region of parameter space in which the mass hierarchy between the mediator and DM is reversed, i.e.\ $m_{Z'} < 2m_{\text{DM}}$. In this regime, the mediator is kinematically forbidden from decaying invisibly into DM pairs. Consequently, DM production proceeds via an off-shell mediator, leading to a suppressed rate compared to the conventional on-shell scenario. Meanwhile, the mediator predominantly decays into visible final states, offering distinctive signatures for experimental searches in the visible sector.

Motivated by this complementary and relatively unexplored phenomenology, we propose a theoretical framework for a light spin-1 DM candidate. In this construction, the SM gauge group is extended by a new non-Abelian gauge symmetry, $SU(2)_D$\footnote{In contrast to previous studies that introduce an additional $SU(2)\times U(1)$ gauge structure \cite{Choi:2019zeb,Ramos:2021txu}, the present framework assumes only a single dark $SU(2)_D$ gauge symmetry, with an Abelian $U(1)_D$ arising solely as a residual subgroup after spontaneous symmetry breaking.}, which breaks spontaneously via the vacuum expectation values (vevs) of two scalar fields charged under the $SU(2)_D$: a triplet, $\Sigma_D$, and a doublet, $\Phi_D$. The corresponding massive gauge bosons form a triplet, comprising the DM particle $X^{\pm}_\mu$, and the mediator $Z'_\mu$. We note that a related vector dark matter scenario has recently been explored in \cite{Foguel:2025hio}. Although both frameworks feature stable vector states arising from broken dark gauge symmetries, the underlying constructions differ, making the approaches complementary.

To connect the dark and visible sectors, we introduce a novel dimension-5 portal operator that embodies a non-Abelian \textit{avatar} of kinetic mixing, that mediates interactions between the DM, the $Z'$, and the SM photon and $Z$ boson. This framework naturally realizes an inverse mass hierarchy between the $Z'$ and the DM, generates momentum-suppressed DM--photon interactions through an effective magnetic dipole operator, and induces additional couplings of the dark scalar fields with the SM. Together, these features open new invisible production channels that can be probed at LDMX and other fixed-target experiments.

Building on the theoretical framework outlined in \secref{section:Model}, we explore the phenomenology of the model in the sub-GeV mass range, considering constraints and signatures across various experimental frontiers. The DM relic abundance is computed under the assumption of thermal freeze-out, as discussed in \secref{section:RelicDensity}. The sensitivity at the direct detection experiments, \textit{viz.}\, DAMIC-M and PANDAX-4T, and projected sensitivities of fixed-target experiments, with particular emphasis on LDMX \cite{LDMX:2025pkp} and NA64 \cite{NA64:2025ddk}, are discussed in \secref{section:direct_det} and \secref{section:FixedTarget}, respectively. In \secref{section:CurrentBounds}, we examine existing constraints on the model parameter space arising from the collider searches, electroweak precision tests, cosmological observations, and astrophysical probes. In \secref{section:Results}, all relevant constraints are placed alongside thermal targets, revealing the complementarity of all bounds and the parameter regions yet to be excluded, followed by concluding remarks in  \secref{section:Conclusion}.

\section{The Model}
\label{section:Model}

We augment the SM gauge group by a new non-Abelian dark gauge symmetry,
\begin{equation}
    SU(2)_L \otimes U(1)_Y \otimes SU(2)_D,
\end{equation}
which contains a scalar doublet ($\Phi_D$) and a scalar triplet ($\Sigma_D$) charged under the $SU(2)_D$. The field content of the dark sector is summarized in Table~\ref{tab:scalar-content}.
\begin{table}[h!]
\centering
\resizebox{\textwidth}{!}{%
\begin{tabular}{c|c|c|c|c}
\hline
Fields & Spin & \textbf{$SU(2)_D$} & Symbols & \textbf{$U(1)_D$} \\
\hline
$\Phi_D$ & 0 & $\mathbf{2}$ & $\Phi_D = 
    \begin{pmatrix}
        \phi_D^+  \\
        \frac{\phi_D^0 + i\chi^0}{\sqrt{2}}
    \end{pmatrix}$ & $\phi_D^\pm: \pm 1/2$ \\
$\Sigma_D$ & 0 &  $\mathbf{3}$ & $\Sigma_D = \frac{1}{2}
    \begin{pmatrix}
        \Sigma_D^0 & \sqrt{2}\Sigma_D^+ \\
        \sqrt{2}\Sigma_D^- & -\Sigma_D^0
    \end{pmatrix}$ & $\Sigma_D^\pm: \pm 1$ \\
$X_\mu^a$ & 1 &  $\mathbf{3}$ & $X^a_\mu =
    \begin{pmatrix}
        X^+_\mu \\
        X^-_\mu \\
        X^0_\mu
    \end{pmatrix}$, $X_\mu^\pm = \frac{1}{\sqrt{2}} (X_\mu^1 \mp X_\mu^2)$ & $X_\mu^\pm: \pm 1$ \\
\hline
\end{tabular}
}
\caption{Particle content and quantum numbers of the dark sector fields.}
\label{tab:scalar-content}
\end{table}
The Lagrangian of the dark sector is given by
\begin{equation}
    \mathcal{L}_{\text{DM}} = 
    - \frac{1}{4} X^{a}_{\mu \nu} X^{a \mu \nu}
    + |D_\mu \Phi_D|^2 + |D_\mu \Sigma_D|^2
    - V(\Phi_D,\Sigma_D),
\end{equation}
where the covariant derivatives are  $D_\mu \Phi_D = \partial_\mu \Phi_D - i \frac{g_D}{2} \tau^a X^a_\mu \Phi_D$, and $D_\mu \Sigma_D = \partial_\mu \Sigma_D - i \frac{g_D}{2} [\tau^a, \Sigma_D] X^a_\mu$, and the scalar potential reads
\begin{align}
    \nonumber
    V(\Phi_D,\Sigma_D) &=  -\mu_{\Phi_D}^2 \Phi_D^\dag \Phi_D 
    - \mu_{\Sigma_D}^2 \text{Tr}[\Sigma_D^2]
    + \mu_{\Phi_D \Sigma_D} \Phi_D^\dag \Sigma_D \Phi_D  \\ 
    &\quad + \lambda_{\Phi_D} (\Phi_D^\dag \Phi_D)^2 
    + \lambda_{\Sigma_D} \text{Tr}[\Sigma_D^2]^2 
    + \lambda_{\Phi_D \Sigma_D} \Phi_D^\dag \Phi_D \text{Tr}[\Sigma_D^2].
\end{align}
The $SU(2)_D$ symmetry is spontaneously broken when the triplet and doublet scalars acquire vevs,
\begin{equation}
    \langle \Sigma_D \rangle = \frac{v_{\Sigma}}{2}
    \begin{pmatrix}
        1 & 0 \\ 0 & -1
    \end{pmatrix},
    \qquad
    \langle \Phi_D \rangle = 
    \frac{1}{\sqrt{2}}
    \begin{pmatrix}
        0 \\ v_{\Phi}
    \end{pmatrix}.
\end{equation}
After spontaneous breaking of the dark gauge symmetry, the masses of the dark vector bosons are obtained as  
\begin{equation}
    m_{X^\pm}^2 = \frac{g_D^2v_D^2}{4},
    \qquad
    m_{X^0}^2 = \frac{g_D^2v_D^2}{4}\sin^2\beta,
\end{equation}
where we define
\begin{align}
    v_D\equiv \sqrt{v_{\Phi}^2 + 2v_{\Sigma}^2}\,, \quad {\rm and} \quad \tan\beta\equiv \frac{v_\Phi}{\sqrt{2}v_\Sigma}\,.
\end{align}
The triplet vev contributes solely to the masses of $X^\pm_\mu$, while the doublet vev generates masses for both $X^\pm_\mu$ and $X^0_\mu$. This distinction arises because $X^\pm_\mu$ carry charge under a global $U(1)_D$ symmetry associated with the $\tau^3$ generator of $SU(2)_D$. The $U(1)_D$ charges of the fields are summarized in the last column of Table~\ref{tab:scalar-content}. The doublet vev spontaneously breaks this $U(1)_D$, leaving behind a residual discrete $\mathbb{Z}_2$ parity that acts as a stabilizing symmetry for the DM. The remnant discrete $\mathbb{Z}_2$ parity of the dark sector acts on the dark gauge bosons as
\begin{align}
    X^{\pm}_\mu \to -X^{\pm}_\mu, \qquad X^0_\mu \to X^0_\mu\,,    
\end{align}
and on the dark scalar fields as
\begin{align}
  \Sigma_D^\pm \to -\Sigma_D^\pm,\quad \Sigma_D^0 \to \Sigma_D^0\,, \quad
\phi_D^\pm \to -\phi_D^\pm,\quad \phi_D^0 \to \phi_D^0, \quad \chi^0 \to \chi^0\,.
\end{align}
This transformation preserves the $SU(2)_D$ algebra and therefore defines a discrete automorphism symmetry of the dark gauge sector, defined by
\begin{align}
PT^{1,2}P^{-1} = -T^{1,2}\,, \quad {\rm and} \quad PT^{3}P^{-1} = T^{3}\,, \quad {\rm where} \quad P=\left(\begin{array}{cc}
    -1 & 0 \\
    0 & 1
\end{array}\right).
\end{align}
Notably, the vacuum configuration remains invariant under this symmetry, so the $\mathbb{Z}_2$ parity survives after spontaneous symmetry breaking.
As a consequence, all interaction vertices contain an even number of ($\mathbb{Z}_2$)-odd fields, implying that the lightest odd particle is stable. We identify the dark-charged gauge bosons $X^\pm_\mu$ as the lightest states carrying nontrivial $\mathbb{Z}_2$ charge, and therefore as the stable DM candidates in this framework.
It is worth noting that this framework naturally realizes an inverse mass hierarchy between the DM and the neutral gauge boson $X^0_\mu$, a distinctive feature that gives rise to interesting phenomenological consequences, as elaborated in the subsequent sections. 

The physical scalar spectrum involves a dark-charged scalar and two neutral scalars. Notably, we neglect the Higgs portal interaction terms in the scalar potential, since the thermal freeze-out of DM via Higgs portal receives stringent bounds from multiple observables, such as Higgs invisible decays, modifications to the triple-Higgs coupling, and di-Higgs production channels \cite{Biekotter:2022ckj,ATLAS:2024ish,CMS:2024awa}. The mass matrices and mixing angles are provided in the \appref{section:appendix_model}.

Here instead, we propose a novel interaction between the dark and visible sectors arising from a dimension-5 non-Abelian gauge kinetic portal operator,
\begin{equation}
    \mathcal{L}_{\text{int}} = -\frac{\sin \zeta}{2v_{\Sigma}} 
    \text{Tr}\left[\Sigma_D X_{\mu \nu}\right] B^{\mu \nu},
    \label{eq:portal}
\end{equation}
where $X_{\mu\nu}\equiv \sigma^a X^a_{\mu\nu}$, and $B_{\mu\nu}$ is the hypercharge field strength tensor and $\zeta$ parametrizes the strength of the interaction, adapting the notation from \cite{Choi:2019zeb}. 

The above  dimension-5 operator can be generated from a fully renormalizable UV complete model by integrating out a heavy vector-like fermion ($\Psi$) charged under both $SU(2)_D$ and $U(1)_Y$ as $(\mathbf{2}, Y_\Psi)$, and with a Yukawa coupling to the triplet scalar $\Sigma_D$ of the form $y\bar\Psi\Sigma_D\Psi$ \cite{Alonso-Alvarez:2023rjq}. At one loop, such states generate the operator $\text{Tr}[\Sigma_D X_{\mu\nu}] B^{\mu\nu}$. The relation between the UV scale, given by the mass of the heavy fermion ($M_\Psi$), and the coefficient of the operator in Eq.~\eqref{eq:portal} is
\begin{align}
    \frac{\sin\zeta}{v_\Sigma} \sim \frac{Y_\Psi g_D g_Y\, y}{16\pi^2 M_\Psi}\,, \qquad {\rm which~implies,} \qquad
M_\Psi \sim \frac{\sqrt{2}\, Y_\Psi g_Y\, y}{16\pi^2} \;\frac{m_{\rm DM} \cos\beta}{\sin\zeta}.
\end{align}
Note that the dependence on $g_D$ cancels in $M_\Psi$, so the validity of effective low energy theory is largely insensitive to the choices of $g_D$. The effective description breaks down when $M_\Psi/Y_\Psi \sim m_{\rm DM}$, i.e.\ for larger values of $\sin\zeta$ or very small couplings $y$, where the mediator cannot be integrated out. For this analysis, we will consider the values of $\sin\zeta$ such that, $M_\Psi$ remains parametrically above the dark sector scale, justifying the effective theory treatment.

When $\Sigma_D$ receives vev, a kinetic mixing between the $X^0_\mu$ and the hypercharge gauge boson is generated as
\begin{equation}
    \mathcal{L}_{\text{mix}} = -\frac{\sin \zeta}{2} 
    X^0_{\mu \nu} B^{\mu \nu},
\end{equation}
which leads to the mixing of $X^0_\mu$ with the SM neutral gauge bosons. We label the  resulting BSM mass eigenstate as $Z^\prime$ (see \appref{section:appendix_model} for further details). In addition to the kinetic mixing, $\mathcal{L}_{\text{int}}$ induces contact interactions of the DM with the $Z$ boson and photon via magnetic dipole operators, as given by,
\begin{align}
    \mathcal{L}_{\rm mag}= i C_{VXX} \partial^{[\mu} V^{\nu]} \left( X^-_\mu X^+_\nu - X^+_\mu X^-_\nu \right)\,,
    \label{eq:AXX}
\end{align}
where
\begin{align}
    C_{VXX} = g_D \sin\zeta \begin{cases} \cos\theta_w\,, &  V_\mu = A_\mu\,, \\
    \left( \sin\theta_w \cos\rho + \tan\zeta \sin\rho \right)\,, & V_\mu = Z_\mu\,, \\
    -\left( \sin\theta_w \sin\rho + \tan\zeta \cos\rho \right)\,, & V_\mu = Z^\prime_\mu\,. \end{cases}   
\end{align}
The parameter $\rho$ is defined in \eqref{def:rho}.

The interactions between the dark scalars,  $\Sigma_D^0$ $\Sigma_D^{\pm}$, and $\Phi_D^0$, and the dark gauge bosons arise directly from gauge invariance and from the dimension-5 non-Abelian kinetic portal introduced in \eqnref{eq:portal}. The resulting feynman rules are written in \appref{section:FeynmanRules}.

The above interactions play the key role to determine the phenomenology of the DM. For our analysis, we adopt a representative benchmark point with 
$\cos\beta = \sin\theta_t = 0.7$, where $\theta_t$ denotes the mixing angle between the neutral scalar states given in \eqnref{eq:theta_t}. The masses of the dark scalars are chosen to lie close to the DM mass scale, 
$m_{\Sigma_D^{\pm}} = m_{\Sigma_D^{0}} = m_{\Phi_D^{0}} = 1.2\, m_{\text{DM}}$, but remain sufficiently heavy to kinematically forbid DM decay into $\Sigma_D^\pm$.

\section{Freeze-Out Relic Density}
\label{section:RelicDensity}

The relic density of DM, measured by \textit{Planck}~\cite{Planck:2018vyg} to be $\Omega_{\text{DM,obs}}h^2 = 0.120 \pm 0.001$, is enforced in our analysis in the so-called \textit{relic targets}, i.e.\ the contours in the parameter space that reproduce this observed value. The Boltzmann equation governing the cosmological DM number density, $n_{\text{DM}}$, is given by,
\begin{equation}
    \dot{n}_{\text{DM}} + 3H n_{\text{DM}} = -\frac{1}{2}\langle \sigma v_{\rm rel} \rangle\, (n_{\text{DM}} ^2 - n_{\text{DM},{\rm eq}}^2) ,
    \label{eq:Boltzmann}
\end{equation}
where $H$ is the Hubble rate, and $\langle \sigma v_{\rm rel} \rangle$ is the thermally averaged cross section of DM annihilation. We have implemented the model in \feynrules \cite{Alloul:2013bka}, computed the thermally averaged cross sections, and solved the Boltzmann equation under the freeze-out approximation \cite{Gondolo:1990dk} to calculate the relic density in terms of the model parameters. We use an in-house developed Boltzmann solver in conjunction with \micromegas \cite{Belanger:2013oya} computed thermally averaged cross sections. For sub-GeV DM, annihilations into hadronic final states become important since below the QCD confinement temperature, quarks no longer exist as free particles~\cite{Izaguirre:2015zva,Ilten:2018crw,PDG}. We therefore incorporate these hadronic channels when determining the relic targets, where they appear as resonances at DM masses satisfying \(m_{\text{DM}} \simeq m_H/2\), with \(m_H\) the mass of the corresponding hadron.

\begin{figure}
\centering
\includegraphics[width=0.5\linewidth]{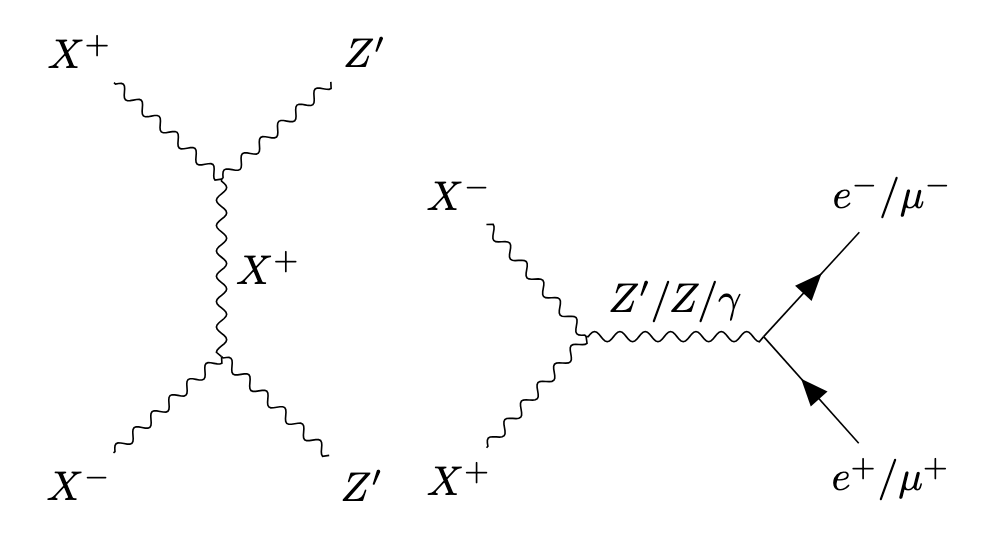}
\caption{DM annihilation processes that set the relic abundance. The left diagram is dominant for small $\zeta$, and the right for $\zeta$ close to 1.}
\label{fig:DMannihilation_diagrams}
\end{figure}

In this model, DM may annihilate into either dark sector or SM particles, with the dominant channels determined by the parameter regime. The thermally averaged cross sections for each process are... \secref{section:relic_density_processes} For small values of $\zeta$, the model enters the \textit{secluded annihilation} regime~\cite{Pospelov:2007mp}, in which DM predominantly annihilates into dark sector states. In this case, the process shown on the left of Fig.~\ref{fig:DMannihilation_diagrams} controls the relic abundance. Additional channels, such as 
$X^+ X^- \to Z'\, \Sigma_D^{0}$,  
$X^+ X^- \to Z'\, \Phi_D^{0}$,  
and co-annihilations like  
$X^{+}\Sigma_D^{-} \to Z'\Phi_D^{0}$  
and  
$X^{+}\Sigma_D^{-} \to Z'Z'$,  
are also included, though they contribute less than 10\% to the total thermally averaged cross section $\langle \sigma v_{\rm rel} \rangle$.

For larger values of $\zeta$ ($\sin\zeta \gtrsim 10^{-2}$), the model transitions to the \textit{direct annihilation} regime, where DM primarily annihilates into SM final states. In this case, channels such as  
$X^+ X^- \to e^+ e^-$,  
$\mu^+ \mu^-$,  
or hadronic states determine the relic density, with the co-annihilation process $X^+ \Sigma_D^- \to \text{SM}$ providing a subdominant contribution.

In the freeze-out scenario, the DM must remain in thermal equilibrium with the SM plasma prior to the decoupling~\cite{Gondolo:1990dk}. Chemical equilibrium is maintained through annihilation (production) into (from) SM species, while kinetic equilibrium is achieved by elastic scattering with SM fermions and with the $Z'$. The corresponding interaction rate,
$R \equiv n\,\langle \sigma v \rangle$, must exceed the Hubble expansion rate, $H$, at temperatures above freeze-out. 

In this work, we explore values of $\sin\zeta$ spanning $10^{-10}$ to $1$. Although this range includes extremely small portal couplings, the relatively large dark gauge coupling, \(g_D \sim 10^{-4} - 10^{-2}\), ensures that the condition $R > H$ is always satisfied. The validity of the standard Boltzmann equation in Eq.~\eqref{eq:Boltzmann} further requires DM to remain in kinetic equilibrium with the SM bath throughout freeze-out~\cite{Binder:2017rgn}; this criterion is also fulfilled across the parameter space considered here.

DM production through freeze-in \cite{Hall:2009bx,Cosme:2021baj}, rather than freeze-out, would occur for small values of $g_D$ -- which we leave as an interesting direction for future work.

\section{Direct Detection at DAMIC-M and PANDAX-4T}
\label{section:direct_det}

\begin{figure}
\centering
\includegraphics[width=0.75\linewidth]{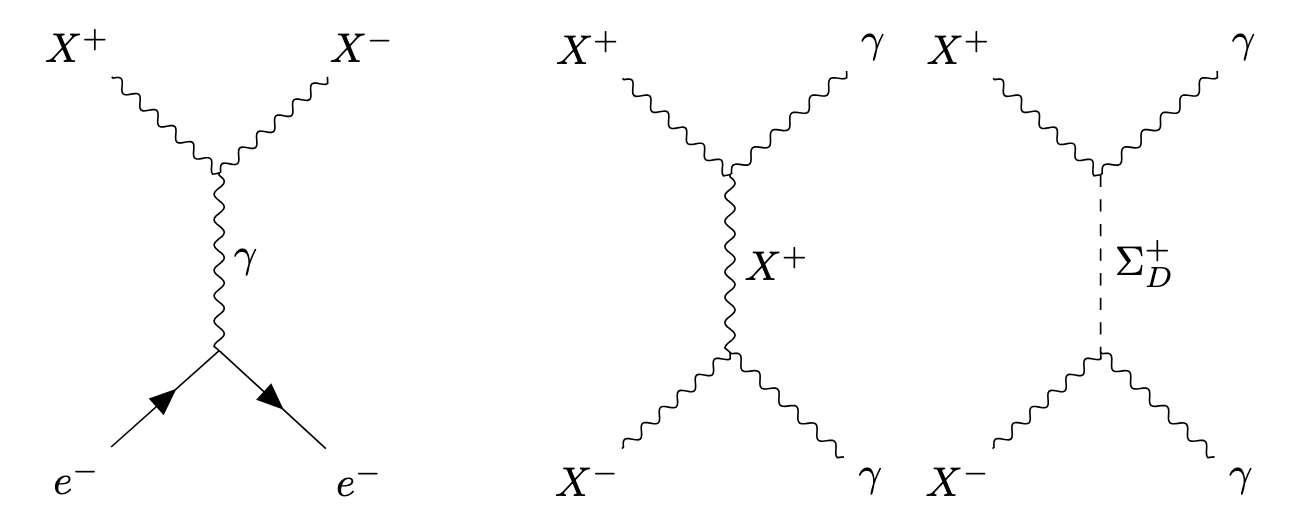}
\caption{Direct and indirect detection processes. Left: DM-electron scattering that induces electronic excitations at direct detection experiments. Right: DM annihilation that inject energy into the CMB through velocity independent thermally averaged cross sections.}
\label{fig:DD_feynman}
\end{figure}

Direct detection experiments serve as the only way to directly measure and discover the DM present in our galaxy \cite{Schumann:2019eaa}\footnote{DM collider and indirect searches act as validation of potential direct detection discoveries \cite{Catena:2024yed}, and as invaluable guiding tools for DM and other new physics beyond the SM.}. For WIMP scale DM, nuclear recoils are the conventional search strategy \cite{Billard:2021uyg}. On the other hand, sub-GeV DM, due to its light mass, would not induce a sufficiently sized nuclear recoil to be observed. However, light DM can induce sizeable electronic excitations in the detector material from DM-electron scattering \cite{Mitridate:2022tnv,Catena:2024rym}. DM-electron scattering proceeds through a t-channel diagram via a virtual photon, as drawn in \figref{fig:DD_feynman}. DM could also scatter via a virtual $Z'$ or $Z$-boson, however we expect these contributions to be sub-leading since the momentum transfer squared is $\ll m_{Z'}^2, m_Z^2$.

From Eq.~\ref{eq:AXX}, we obtain the following non-relativistic amplitude for DM–electron scattering
\begin{align}
i \mathcal{M}= -
\frac{i e g_D \sin\zeta \cos\theta_w}{|\boldsymbol{q}|^2} &\Bigg[
    2 \frac{m_e}{m_{\rm DM}}\delta^{s's}\boldsymbol{q} \cdot \boldsymbol{\mathcal{S}}_{\rm DM}^{\lambda'\lambda}\cdot \boldsymbol{q}
    + 2 m_e \delta^{s's} \left(i\boldsymbol{q} \times \boldsymbol{v}^\perp \right) \cdot \boldsymbol{S}_{\rm DM}^{\lambda'\lambda}
    \nonumber\\
    & + 2|\boldsymbol{q}|^2  \boldsymbol{S}_{e}^{s's} \cdot \boldsymbol{S}_{\rm DM}^{\lambda'\lambda}
    -2 \left( \boldsymbol{q} \cdot \boldsymbol{S}_{e}^{s's} \right)\left( \boldsymbol{q} \cdot \boldsymbol{S}_{\rm DM}^{\lambda' \lambda} \right)
     \Bigg]\,.\nonumber\\
    \label{eq:MNR}
\end{align}
Here, we adopt the notation of~Ref.~\cite{Catena:2019gfa} where $m_e$ is the electron mass, $\boldsymbol{q}$ is the momentum transfer, $\boldsymbol{v}^\perp$ is the transverse DM-electron relative velocity while $\boldsymbol{S}_e^{s's}=\xi^{\dagger s'}\boldsymbol{S}_e \xi^{s}$, $\boldsymbol{S}_e=\boldsymbol{\sigma}/2$ is the electron spin operator, $\boldsymbol{\sigma}=(\sigma_1,\sigma_2,\sigma_3)$ is a three-dimensional vector whose components are the three Pauli matrices, $\xi^s$, $s=1,2$, is a two-component spinor, $\boldsymbol{S}_{DM}^{\lambda' \lambda} = - i \boldsymbol{e}'_{\lambda'} \times \boldsymbol{e}_{\lambda}$, $\mathcal{S}^{\lambda' \lambda}_{{\rm DM}\, ij} = \frac{1}{2} \left(e_{\lambda i} \, e'_{\lambda' j}+ e_{\lambda j} \, e'_{\lambda' i}\right)$ and $e_{\lambda i}$ is the $i$-th component of the $\lambda$-th DM polarization vector. From Eq.~\ref{eq:MNR} and the null results reported by DAMIC-M \cite{DAMIC-M:2025luv} and PANDAX-4T \cite{Zhang:2025ajc}, we derive 90\% C.L. exclusion limits in the $(m_{\rm DM}, \sin\zeta)$ plane. For DAMIC-M, we perform this calculation by requiring that the probability of observing a total number of signal plus background events smaller than or equal to the actually observed one in the one, two and three electron channels is at least 0.1. In doing so, we use the expected background reported in Tab.~1 of \cite{DAMIC-M:2025luv}, and compute the rate of DM-induced electronic excitations in a silicon detector using the crystal response functions obtained in \cite{Catena:2021qsr}. We assume the total detector efficiency given in \cite{DAMIC-M:2025luv} and an exposure of 1.257 kg-year. For PANDAX-4T, we impose that, in each of the eight ionization-charge bins considered in Ref.~\cite{Zhang:2025ajc}, the probability of observing a number of signal plus background events smaller than or equal to the actually observed one is at least 0.1. We extract the PANDAX-4T background model from Fig.~2 of Ref. \cite{Zhang:2025ajc} and compute the differential electron recoil energy rate using the xenon response functions derived in Ref. \cite{Catena:2019gfa}. We then convert this into an ionization rate by applying the detector response model introduced in Ref. \cite{Essig:2012yx}, the total detector efficiency reported in Fig.~1 of Ref. \cite{Zhang:2025ajc} and employing an exposure of 1.04 ton-year. For both experiments, we assume Poisson statistics for the observed number of counts.

\section{Sensitivity at Fixed-Target Experiments}
\label{section:FixedTarget}

In this section, we consider the most relevant experimental constraints and projections, including constraints from fixed-target experiments, colliders, direct detection, electroweak precision measurements, BBN, and from energy injection into the CMB.

With the ongoing development of future fixed-target experiments, such as LDMX \cite{LDMX:2025pkp,LDMX:2018cma}, and the current competitive limits coming from NA64 \cite{NA64:2025ddk} and electron beam dump experiments \cite{Andreas:2012mt}, fixed-target experiments are capable of extensively probing sub-GeV DM for numerous theoretical scenarios \cite{Catena:2023use,Catena:2025fsl,Berlin:2020uwy,Krnjaic:2022ozp,Gori:2022vri,Cosme:2021baj}.
Having a high energy beam of electrons (or protons in the case of SHiP \cite{SHiP:2020noy} and other proton beam dumps) incident on a stationary target dense with nuclei, allows for a large number of potential DM events that can be detected through their missing energy or momentum (or through the direct scattering with a downstream detector in the case of proton beam dumps). However, for our scenario since $m_{Z'} < m_{\text{DM}}$, production of DM proceeds through off-shell processes and is thus suppressed relative to its on-shell counterpart. The on-shell production of visible species, in particular $e^+ e^-$ also occurs, acting as a complementary probe of this model.
We consider the missing momentum search constraints from current NA64 null results and the sensitivity projections of the future experiment LDMX, in addition to the visible searches from past beam dump experiments E774 \cite{Bross:1989mp}, E141 \cite{Riordan:1987aw}, Orsay \cite{Davier:1989wz}, E137 \cite{Bjorken:1988as}, CHARM \cite{Gninenko:2012eq}, KEK \cite{Konaka:1986cb}, and NuCal \cite{Tsai:2019buq}.

NA64's 100 GeV electron beam incident on an active target composed of lead absorber layers, with $4.4 \times 10^{11}$ electrons on target (EOT) so far, places competitive constraints on DM and dark sector particles that can be produced through these beam target interactions \cite{NA64:2023wbi, NA64:2025ddk}. We consider both visible and invisible production, where for the latter the production of DM through dark bremsstrahlung and through the decay of vector mesons are each capable of delivering missing energy signatures.

LDMX is a future missing-momentum experiment that will operate in two phases, where the second phase plans to have $10^{16}$ EOT with a beam energy of 8 GeV \cite{LDMX:2023zbn,LDMX:2018cma}. With the ability to measure the transverse momentum, in addition to the energy, of the recoil electron after the interaction with the tungsten target, LDMX is subject to less background, projecting sensitivities reaching further than previous experiments in the on-shell case. In this study we explore the less studied and less sensitive off-shell / visible decay regime \cite{Berlin:2018bsc}. Similarly to NA64, LDMX will search for signatures from visible decays, dark bremsstrahlung, and invisible vector meson decay \cite{Schuster:2021mlr}.

\begin{figure}
    \centering
    	 \includegraphics[width=0.6\linewidth]{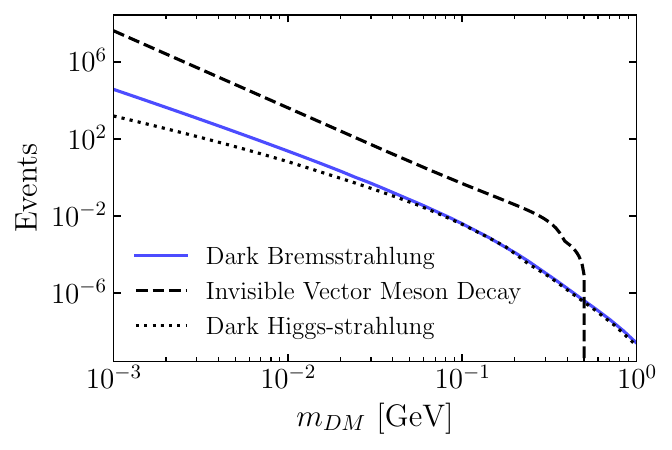}
    	\caption{The number of DM events expected at LDMX Phase II, with a beam energy of 8 GeV and $10^{16}$ EOT, for two DM production channels: dark bremsstrahlung (in solid) and invisible vector meson decay (in dashed). The model parameters are set to: $\zeta = 10^{-3}$, $g_D = 0.01$, and the scalar masses are $m_{\Sigma_D^{\pm}}=m_{\Sigma_D^{0}}=m_{\Phi_D^0} = 1.2 \: m_{\text{DM}}$.}
    \label{fig:Events_bremIVM}
\end{figure}
The two invisible production channels considered here for LDMX and NA64, dark bremsstrahlung (and dark Higgs-strahlung) and invisible vector meson decay, are plotted in \figref{fig:Events_bremIVM} with LDMX experimental parameters and Phase II statistics. The analogous plot for current NA64 parameters and statistics is qualitatively similar, except with less events for both channels. Interestingly, the number of events from invisible vector meson decay exceeds dark bremsstrahlung by roughly five orders of magnitude for DM masses where mesons are kinematically allowed to decay to DM. 
Below, we discuss each channel, invisible and visible, in more detail.

\subsection{Dark Bremsstrahlung}
\begin{figure}
\centering
\includegraphics[width=0.8\linewidth]{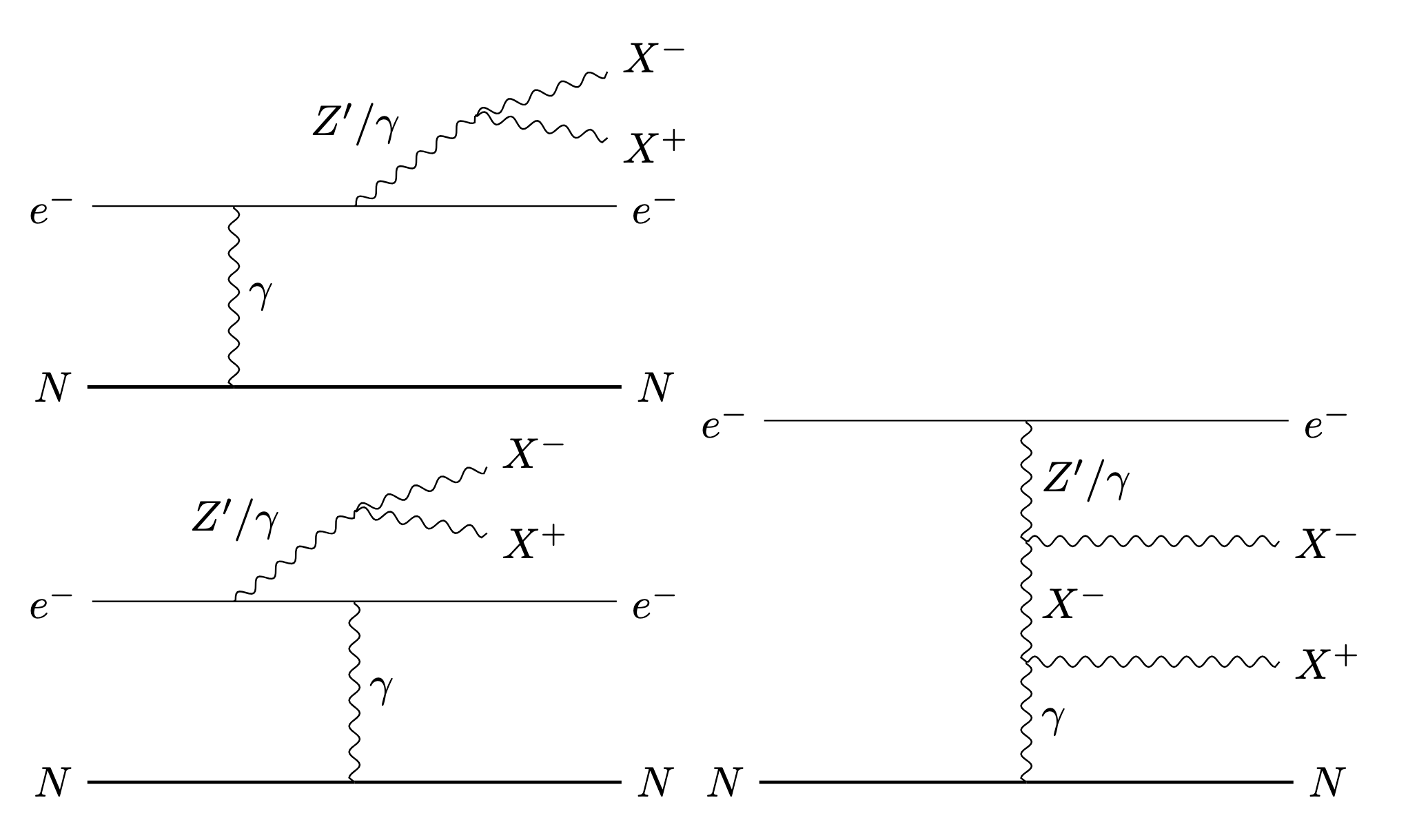}
\caption{The dominant off-shell dark bremsstrahlung processes generating DM that occur at fixed-target experiments, where $N$ represents the target nucleus. Left: DM production diagrams off the initial or final state beam electron. Right: Bethe-Heitler trident diagrams for DM production.}
\label{fig:DarkBrem}
\end{figure}

Beam electrons incident on a target could produce dark sector particles in a process similar to ordinary bremsstrahlung. In the usual dark photon kinetic mixing scenario where $m_{A'} > 2m_{\text{DM}}$ \cite{Berlin:2020uwy}, the dark photon is produced on-shell and later decays to DM, therefore the event rate is independent of the DM properties and is only dependent on the dark photon.
In our scenario however, there is the off-shell production of DM through a virtual $Z'$ or ordinary photon, since the mass hierarchy is reversed such that $m_{\text{DM}} > m_{Z'}$. Off-shell production of DM is suppressed compared to on-shell production \cite{Berlin:2018bsc}, leading to less expected events.
Dark bremsstrahlung processes that occur dominantly for our choice of model parameters, namely the emission of DM off the initial or final state electron, and dark Bethe-Heitler trident processes, are drawn in \figref{fig:DarkBrem}.

Dark bremsstrahlung diagrams involving the new scalars, $\Phi_D^0$, $\Sigma_D^0$, and $\Sigma_D^\pm$, as intermediate states, become relevant as $\sin \beta$ approaches 0, however for the choice of parameters where $\cos \beta = 0.7$, these diagrams contribute negligibly to the production of dark states. 
Simulations of dark bremsstrahlung events are performed in \madgraph\ \cite{Alwall:2014hca}, with UFO files \cite{Degrande:2011ua} generated by our \feynrules\ model file \cite{Alloul:2013bka}. The DM and dark scalar vertices, in addition to the nucleus-photon vertex with form factor from \cite{Bjorken:2009mm}\footnote{The form factor from A18 and A19 of \cite{Bjorken:2009mm}, which is used as an approximate representation of the effects from the atomic/nuclear structure, has a typo in Eq. A19: the second term should not be squared.}, are included in our UFO files.

As evident by \figref{fig:Events_bremIVM}, dark bremsstrahlung events are much less than that of invisible vector meson decay, due to the off-shell nature of the dark brem.

\subsection{Dark Higgs-strahlung}

\begin{figure}
\centering
\includegraphics[width=0.75\linewidth]{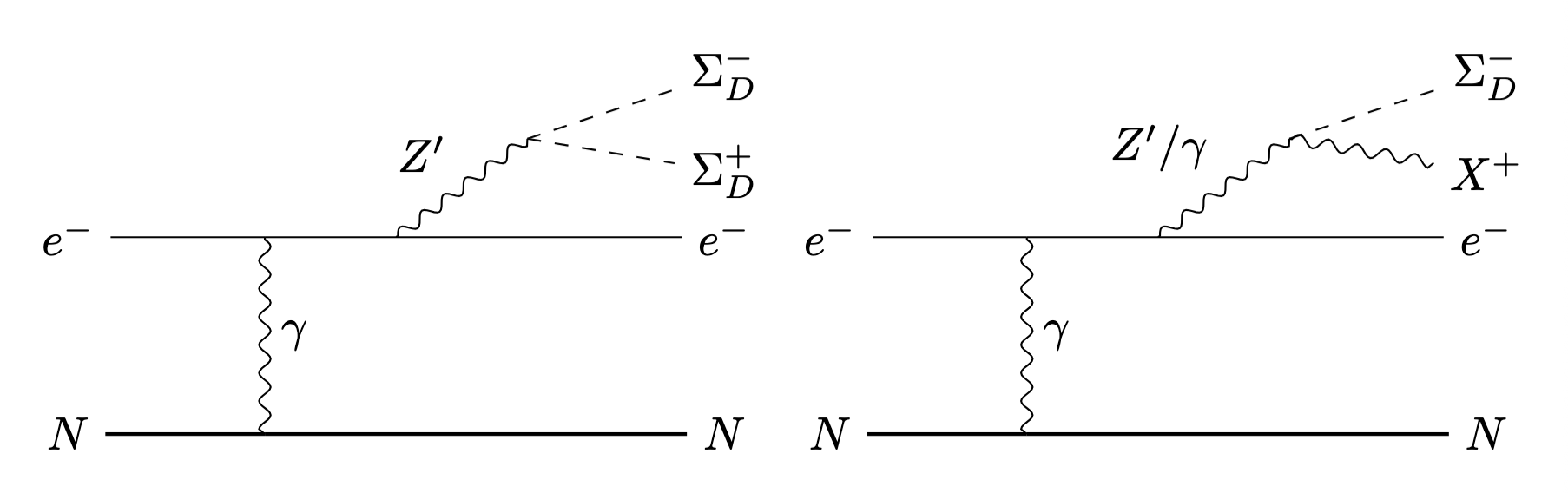}
\caption{Examples of dark Higgs-strahlung processes providing additional missing energy/momentum signatures.}
\label{fig:DarkHiggs_diagrams}
\end{figure}

Not only can the production of DM itself produce a missing momentum signature, but also the production of new scalars \cite{Acanfora:2025ikj,Duerr:2017uap}, since similarly to DM, dark scalars would not interact in the detector. Through the interaction vertices listed in \appref{section:FeynmanRules}, dark scalar production occurs via dark Higgs-strahlung -- analogously to normal dark bremsstrahlung.
In our set up, the production of scalars $\Sigma_D^{\pm / 0}, \Phi_D^0$ through the process $e^- N \to e^- N S $, where $S$ is final states such as $\Sigma_D^{+(-)} X^{-(+)}$, $\Sigma_D^+ \Sigma_D^-$, and $\Sigma_D^0/\Phi_D^0$, would look like a missing energy/momentum signature at an experiment like LDMX, similarly to a signature from DM production. The final states $\Sigma_D^{+(-)}X^{-(+)}$ make up the largest contribution of events. The processes with final states $\Phi_D^0 Z'$ and $\Sigma_D^0 Z'$ also occur, however there is only \textit{missing} energy/momentum if the $Z'$ decays visibly outside of the detector.

We draw the most prominent dark Higgs-strahlung diagrams resulting in missing energy/momentum signatures in \figref{fig:DarkHiggs_diagrams}. Trident style diagrams where the new scalars are produced from the virtual photon exchanged by the nucleus also exist, although we find their contribution to be sub-leading. As evident by \figref{fig:Events_bremIVM}, the dark Higgs-strahlung cross section is of similar magnitude to dark bremsstrahlung.

\subsection{Invisible Vector Meson Decay}

In addition to the dark bremsstrahlung process, DM can also be produced invisibly through vector meson decays generated in beam-target collisions -- the so-called \textit{invisible vector meson decay} \cite{Schuster:2021mlr}. Bremsstrahlung photons are converted to vector mesons through exclusive photoproduction processes, which then decay to DM through a mediator that mixes with vector mesons. The three possible vector mediators in this model are the photon, $Z$ boson, and $Z'$. The $Z$ boson is heavy compared to the vector mesons, resulting in a much smaller cross section than those of the other two mediators. There is a larger cross section through the ordinary photon, however not as high as through the $Z'$ -- which has a mass less than the DM but similar order of magnitude. Therefore, we consider the process $V \to Z' \to X^+ X^-$, where $V$ is a vector meson such as the $\rho$, $\omega$, $\phi$, or $J/\psi$, and the $Z'$ is off-shell.  Since the $Z'$ couples to SM quarks through vector currents, the $Z'$ mixes with the vector mesons with the form factors given in the fourth line of Table III of \cite{Schuster:2021mlr}. Following the methodology of  \cite{Schuster:2021mlr}, and adapting to this model, we compute the predicted number of DM events from invisible vector meson decays for the experiments NA64 \cite{NA64:2025ddk} and LDMX \cite{LDMX:2025pkp,LDMX:2018cma}. The number of events from the decay of a vector meson $V$ is given by,
\begin{equation}
	N_{\text{DM}} =  \sum_V N_V Br(V \to X^+ X^-),
\end{equation}
where $N_V$ is the predicted number of vector mesons produced from beam target collisions and $Br(V \to  X^+ X^-) \equiv \frac{\Gamma_{V \to X^+ X^-}}{\Gamma_V}$ where the decay rate to DM is written in \appref{section:appendix_CrossSections} and $\Gamma_V$ is measured experimentally and reported in PDG \cite{ParticleDataGroup:2018ovx}. The sum runs over \mbox{$V=\rho,\omega,\phi$} and in the case of NA64, additionally $J/\psi$ since the beam energy of 100 GeV is sufficiently high.
 We use the corresponding values found in Table II of \cite{Schuster:2021mlr} for $N_V$, calculated using a combination of experimental measurements and theory. We take the currently reported EOT at NA64 to be $4.4 \times 10^{11}$, and thus scale-up the corresponding values in Table II of \cite{Schuster:2021mlr} to be $N_\rho = 1.1 \times 10^7$, $N_\omega = 7.9 \times 10^5$, $N_\phi = 7.0 \times 10^5$, and $N_{J/\psi} = 9.7 \times 10^3$. For LDMX, we take the corresponding values for Phase II reported in Table II of \cite{Schuster:2021mlr}.
 
    \begin{figure}
    \centering
	\includegraphics[width=0.6\linewidth]{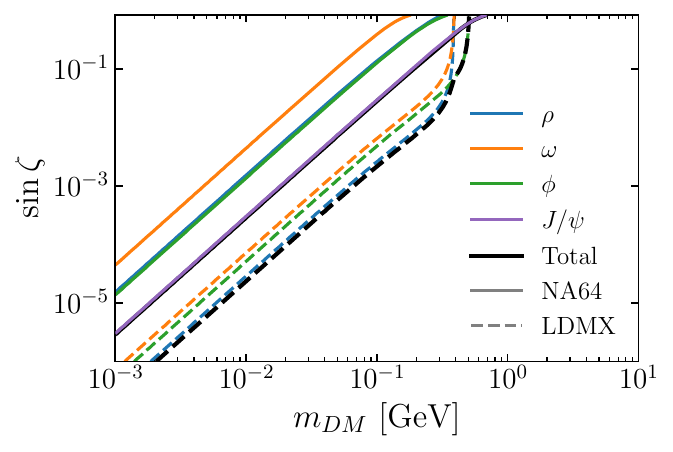}
	\caption{90\% C.L. constraints from current NA64 null results \cite{NA64:2025ddk} (in solid) and projections for LDMX Phase II \cite{LDMX:2025pkp} (in dashed), from invisible vector meson decay.  The contribution from each vector meson is plotted in colour, and the total in black. $g_D$ is set to $10^{-2}$.}
	\label{fig:IVM}
\end{figure}
90\% C.L. constraints and projections for invisible vector meson decays at NA64 and LDMX, respectively, for each specie of meson are plotted in \figref{fig:IVM} as a function of the model parameters. The reach of LDMX surpasses that of NA64 due to the increased statistics, providing a significant improvement to the sensitivity landscape from the data that will come from LDMX. Interestingly, since the $V \to X^+ X^-$ decay rate is proportional to $1/m_{\text{DM}}^4$, the sensitivity increases as $m_{\text{DM}}$ decreases -- resulting in much more sensitivity for lower masses compared to the on-shell dark photon studied in \cite{Schuster:2021mlr}, which features a sensitivity that is independent of $m_{\text{DM}}$. Due to the inverse mass hierarchy, where $m_{\text{DM}} > m_{Z'}$, the sensitivity lacks the resonance that exists in the on-shell dark photon case; since at $m_{Z'} \approx m_V$, the condition $m_V>2m_{\text{DM}}$ for the decay to be allowed, would not hold.

As evident by \figref{fig:Events_bremIVM}, the sensitivity from inverse meson decays exceed that of dark bremsstrahlung.

\subsection{Visible Searches}
Finally, the $Z'$ decays visibly to SM final states $f^+ f^-$, the only decay channel kinematically allowed, and can be searched for as displaced visible energy depositions in the detectors \cite{Berlin:2018bsc,Andreas:2013lya,Gninenko:2013rka, LDMX:2025pkp} or as missing energy/momentum if it decays outside of the detector. We consider displaced visible energy deposition limits from existing experiments NA64 \cite{NA64:2019auh}, electron beam dump experiments E774, E141, Orsay, KEK, and E137 \cite{Andreas:2012mt}, proton beam dumps CHARM and NuCal \cite{Blumlein:2011mv,Tsai:2019buq}. 

For LDMX Phase II projections, we compute limits from displaced visible events and also events from missing energy/momentum in which the $Z'$ decays outside of the detector.
The number of events from a visibly decaying $Z'$ is \cite{Berlin:2018bsc},
\begin{equation}
    N_{\text{vis}} = N_{Z'} \big( e^{-z_{\text{min}}/\gamma c \tau} - e^{-z_{\text{max}}/\gamma c \tau}  \big),
\end{equation}
where $z_{\text{min}}$ and $z_{\text{max}}$ are the distance to the beginning and end of the detector, respectively, $\gamma$ is the $Z'$ boost, and $c \tau$\footnote{We restore $c$ for clarity when quoting decay lengths in physical units.} is the proper decay length of $Z'$.

For LDMX Phase II we take $10^{16}$ EOT, with $z_{\text{min}} = 43$ cm and $z_{\text{max}} = 315$ cm for the visible signature, and $z_{\text{min}} = 315$ cm and $z_{\text{max}} = \infty$ for the invisible signature. We use the approximate expression for the number of $Z'$s produced in beam-target collisions as a function of $m_{Z'}$ and couplings \cite{Berlin:2018bsc}, 
\begin{equation}
    N_{Z'} \approx \frac{7}{10^{-5}} \Big( \frac{ 4 \sin^2\theta_w \sin \rho - \sin \rho + 3 \cos \rho \tan \zeta \sin\theta_w}{4 \cos \theta_w \sin \theta_w } \Big)^2 \Big( \frac{100 \: \text{MeV}}{ m_{Z'}}\Big)^2.
\end{equation}

\section{Other Relevant Bounds}
\label{section:CurrentBounds}

In this section, we provide a brief overview of the constraints on the model parameter space arising from existing terrestrial experiments and astrophysical observations. 

\subsection{Colliders: Direct Search, Electroweak Precision, and Higgs Constraints}

$Z'$s decaying into SM final states, either promptly or long lived, after being produced in proton-proton or $e^+ e^-$ collisions are searched for at collider experiments. We consider the limits from the null results of the experiments LHCb \cite{LHCb:2019vmc}, KLOE \cite{KLOE-2:2018kqf}, BaBar \cite{BaBar:2014zli}, NA48/2 \cite{NA482:2015wmo}, and FASER \cite{FASER:2023tle}, most relevant for the sub-GeV DM mass range.

New physics is capable of affecting SM electroweak precision observables, such as neutral current interactions, leading to constraints on the parameter space of the beyond the SM model. Following \cite{Burgess:1993vc,Tran:2024srm,Alonso-Gonzalez:2025xqg}, we determine the effect of this model as a function of $\sin\zeta$ on the electroweak oblique parameters $S$ and $T$ \cite{Peskin:1991sw}.
The neutral current interaction in our scenario can be split into the SM and the beyond the SM contributions,
\begin{equation}
    \mathcal{L}_\text{nc} = \mathcal{L}_\text{SM,nc} -\frac{e \sin \rho \tan \zeta \sin \theta_w}{\sin \theta_w \cos \theta_w} \sum_i \bar{f_i} \gamma^\mu \Big( -T_{3i}P_L + Q_i \Big) f_i Z_\mu,
\end{equation}
where $Q_i$ is the charge of the corresponding fermion and $T_{3i}$ is the fermion’s third component of weak isospin. We match the above equation to equation 23 of \cite{Burgess:1993vc} in order to write expressions for the $S$ and $T$ parameters in our model.
Using the experimentally determined values from \cite{ParticleDataGroup:2024cfk}, $S=-0.05 \pm 0.07$ and $T=0.00\pm0.06$, we compute the $\chi^2$ as outlined in \cite{Alonso-Gonzalez:2025xqg} and find the following 95\% confidence level (C.L.) upper limit,
\begin{equation}
    \sin \zeta \leq 0.02.
\end{equation}
We find that these constraints are not competitive with the other constraints considered in this study. Furthermore, the predicted shifts in the $Z$-boson mass are sufficiently small that they do not impose additional meaningful bounds on the parameter space.
We leave a more comprehensive analysis of the loop-induced electroweak contributions arising from the extended scalar sector, which may further modify the $S$ and $T$ parameters in addition to the Z-boson mass, to future work.

The invisible branching ratio of Higgs decays are excluded above 0.145 at 95\% C.L. \cite{ATLAS:2022yvh}. In this scenario, the Higgs can decay invisibly through the process $H \to Z' Z'$, with a decay rate given in \eqnref{eq:InvHiggs}. The branching ratio of this process is small for parameter values that we consider, with a value of $\text{BR}_{H\to Z' Z'} \approx 4.6 \times 10^{-13}$ for $\sin \zeta = 0.1$ and $m_{\text{DM}} = 1$ GeV. Therefore, we remain safe from invisible Higgs decay constraints.

\subsection{Astrophysical and Cosmological Constraints}

\subsubsection*{Energy Injection into the CMB}
The annihilation of DM particles in the early universe injects energy into the CMB, which is constrained by current observations since it can change the recombination history, thus modifying the measured temperature and polarization power spectra \cite{Planck:2015fie}.  For constraints from the CMB to be relevant, DM must annihilate with a cross section that is s-wave, such that it is not velocity suppressed\footnote{There does indeed exist bounds on p-wave annihilating DM \cite{Liu:2016cnk}, however they impose constraints that are not competitive with the other constraints we already consider in this work.}. In this model, DM annihilation into photons through the diagrams displayed in \figref{fig:DD_feynman} are s-wave dominant, therefore the cross section is constant at low temperatures. We thus apply the 95\% C.L. bound derived in  \cite{Slatyer:2015jla} on the channel $\text{DM DM} \to \gamma \gamma$.

\subsubsection*{Big Bang Nucleosynthesis}

Dark sector particles, especially of sub-GeV mass, thermally coupled to the SM and present during Big Bang Nucleosynthesis (BBN) can lead to alterations in the number of relativistic degrees of freedom, $N_{eff}$, affecting primordial abundances of light elements.
The abundances of light elements are measured to a given precision, and any expected deviations from these observations lead to constraints on dark sector models \cite{Giovanetti:2021izc,Sabti:2019mhn,Depta:2019lbe}. A detailed consideration of constraints from BBN is beyond the scope of this work, however we include a conservative lower bound on the mass of the lightest dark species, $Z'$, to be $> 10$ MeV. Thanks to the large dark coupling values, $g_D = 10^{-4}, 10^{-3}, 10^{-2}$, the $Z'$ remains in thermal equilibrium with the SM thermal bath for all values of $\sin \zeta$ (down to $\sin \zeta = 10^{-10}$), therefore $m_{Z'}>10$ MeV applies across the entire parameter space considered here.

\subsubsection*{Indirect Detection}
Searches of DM annihilation products in astrophysical environments constrain the annihilation cross section of DM \cite{Cirelli:2025rky}. DM annihilation into SM fermion final states such as $e^+ e^-$, $q \bar{q}$, proceed via p-wave (velocity) suppressed cross sections, leaving constraints through these channels weak. We thus only consider the velocity independent annihilation into $\gamma \gamma$ channel, drawn in \figref{fig:DD_feynman} (right), where the dominant constraints come from CMB as discussed above.

\subsubsection*{Bullet Cluster}
 DM self-interactions are constrained by measurements of the colliding galaxy cluster, the Bullet Cluster \cite{Markevitch:2003at,Wittman:2017gxn}. 
The cross section of DM self-interactions at non-relativistic energies, for all parameter values considered in this study, fall safely below the upper limit of $\sigma/m_{\text{DM}} < 1 \: \text{cm}/\text{g}$ reported in \cite{Markevitch:2003at}.

\section{Relic Targets, Constraints, and LDMX Projections}
\label{section:Results}
\begin{figure}[h]
	\centering
	\includegraphics[width=\linewidth]{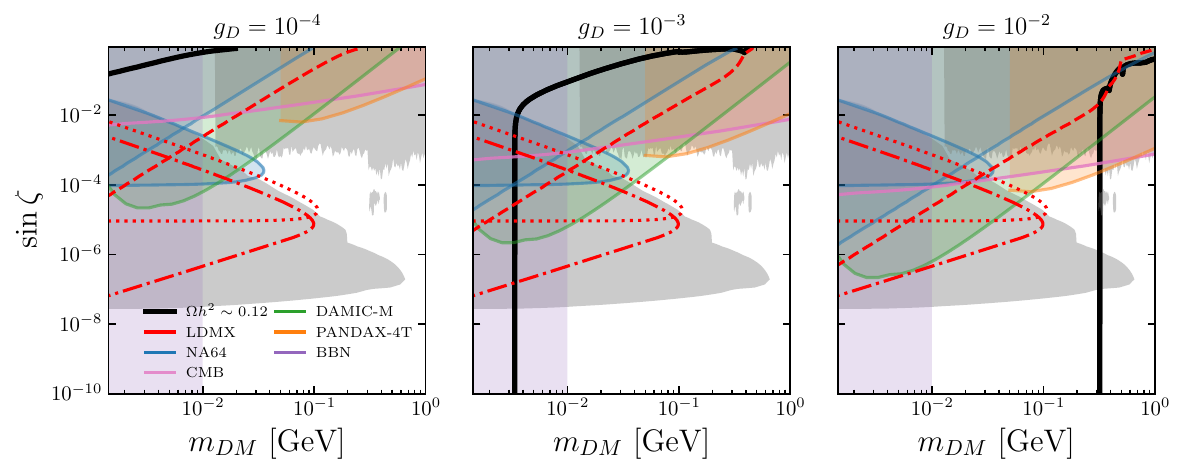}
	\caption{Summary of the most relevant constraints on this model in the $\sin \zeta$ vs $m_{\text{DM}}$ parameter space, overlaid with the relic targets (curves reproducing the observed relic abundance consistent with Planck) \cite{Planck:2018vyg} shown as thick solid lines. The three subplots differ by the value of the $g_D$ parameter: with $g_D = 10^{-4}$ on the left, $g_D=10^{-3}$ in the middle, and $g_D = 10^{-2}$ on the right. 90\% C.L. projections for LDMX Phase II \cite{LDMX:2025pkp} are drawn in red: invisible vector meson decay is drawn in dashed, and visible decay of the $Z' \to e^+ e^-$ outside of the detector in dash-dot and inside of the detector in dotted.
    90\% C.L.s for NA64 \cite{NA64:2025ddk} from invisible vector meson decay and visible $Z'$ decay \cite{NA64:2018lsq} at current statistics are in blue. The pink curves indicate regions in which the parameters are constrained by CMB measurements \cite{Slatyer:2015jla} at 95\% C.L.. Direct detection 90\% C.L. constraints from DAMIC-M are plotted in green \cite{DAMIC-M:2025luv} and from PANDAX-4T in orange \cite{PandaX:2024qfu}. Grey regions are excluded at 95\% C.L. by the beam dump experiments CHARM, E774, E141, Orsay, KEK, and E137 \cite{Andreas:2012mt}, and at 90\% by NuCal \cite{Tsai:2019buq}, along with collider experiments LHCb \cite{LHCb:2019vmc}, KLOE \cite{KLOE-2:2018kqf}, BaBar \cite{BaBar:2014zli}, NA48/2 \cite{NA482:2015wmo}, and
    FASER \cite{FASER:2023tle} at 90\%.} 
\label{fig:Sensitivity}
\end{figure}

\figref{fig:Sensitivity} presents the leading constraints on our model in the $\sin \zeta$ vs $m_{\text{DM}}$ plane, for three benchmark values of $g_D$.
Constraints from current experimental data are marked as shaded regions, while projected constraints from the up-coming LDMX are in dashed, dotted, and dash-dotted, corresponding to invisible vector meson decay, visible decays, and visible decays outside of the detector, respectively. 

The relic targets in black vary depending on the \textit{regime} defined by the parameter values and corresponding dominant DM annihilation processes in the early universe. The \textit{secluded} regime, as discussed in \secref{section:RelicDensity}, occurs for values of $\sin \zeta \lesssim 0.01$. In this regime, only dark sector processes (left diagram of \figref{fig:DMannihilation_diagrams}) contribute to the relic abundance, therefore the relic target is independent of $\zeta$. The cross section of $X^+ X^- \to Z' Z'$ is inversely proportional to $m_\text{DM}$, thus increasing the coupling results in the relic target shifting to larger DM masses, explaining the shift to larger masses with increasing $g_D$. One can imagine how values of $g_D$ in between or outside of the values considered would alter the relic targets. If $g_D = 0.1$, we are no longer in the sub-GeV DM regime, therefore we leave that investigation for future work.
The \textit{direct annihilation} regime occurs for $\sin \zeta \gtrsim 0.01$, where DM annihilation to SM final states sets the relic abundance. This regime is dependent on $\zeta$, thus exhibits a sloped relic target with a kink associated with the onset of annihilations to $\mu^+ \mu^-$, and resonances corresponding to hadronic final states.

Constraints from the null result of existing beam dump and collider experiments searching for visible signatures from dark sectors, indicated by grey shaded regions, provide large coverage over the parameter space. 
Interestingly, for both LDMX and NA64, the sensitivity from invisible meson decays surpasses that of dark bremsstrahlung (including dark Higgs-strahlung), as discussed in more detail in \secref{section:FixedTarget}.  Limits from the CMB extend the constrained parameter region for $g_D = 10^{-2}$, where they slightly outperform existing collider constraints. Although projections of LDMX Phase II (in red), for both visible and invisible searches, do not dramatically extend the sensitivity, it will however provide valuable validation of previous results, such as those from FASER. Furthermore, direct detection from DAMIC-M and PANDAX-4T, in green and orange respectively, provide even stronger constraints than what LDMX is expected to provide.
These results are not unexpected; being in the kinematic regime where the mediator cannot be produced on-shell leads to off-shell suppression of invisible final states, and the availability of visible search channels.
Notice how all constraints that are dependent on the dark sector coupling, $g_D$, (direct detection, CMB, and invisible vector meson decays at fixed-targets), become stronger compared to visible searches, as $g_D$ becomes larger. These results highlight the complementarity between different types of DM searches -- from cosmological, laboratory, and astrophysical, searches.

Evidently, $g_D \lesssim 10^{-3}$ is ruled out by BBN and current experimental search limits. This leaves $g_D \sim 10^{-2}$, for $\sin \zeta \lesssim 4 \times 10^{-8}$ down to arbitrarily small values, in addition to the gap in sensitivity around $\sin \zeta \sim 10^{-5}$ for $m_\text{DM} \gtrsim 100$ MeV, still viable.

\section{Conclusion}
\label{section:Conclusion}
We consider a viable light spin-1 DM candidate that interacts with the SM through a dimension-5 portal that gives rise to off-shell invisible and visible production at fixed-target experiments such as NA64 and the future LDMX. 
Specifically, we explore a model where the SM gauge group is augmented by a non-Abelian dark $SU(2)_D$ symmetry, which is spontaneously broken by the vacuum expectation values of a scalar doublet and triplet. The extended scalar sector that we outline, is capable of explaining the mass generation of DM and the $Z'$, and also plays a role phenomenologically at fixed-target experiments. This framework naturally realizes an inverse mass hierarchy where the DM is heavier than the $Z'$.

This mass relationship leads to a distinctive phenomenological landscape. The mediator is kinematically forbidden from decaying invisibly into DM pairs, therefore DM  production at fixed-target experiments giving rise to missing energy/momentum signatures must proceed through suppressed off-shell processes: dark off-shell bremsstrahlung and invisible vector meson decay. In contrast to the on-shell case, where the mediator can be resonantly produced and subsequently decay into invisible final states yielding large cross sections, the off-shell regime suppresses all such rates by additional powers of the mediator propagator, fundamentally reshaping the hierarchy of experimental sensitivities. While these missing energy/momentum signatures are usually the dominant probe in on-shell scenarios, they fall short of other probes such as direct detection in off-shell scenarios. In this work, we quantified the interplay between these different frontiers, demonstrating that direct detection provides a dominant probe in this regime where fixed-target signatures are suppressed by off-shell kinematics.
Direct detection experiments DAMIC-M and PANDAX-4T are found to be the current dominant constraints for this model, providing critical sensitivity that complements and often surpasses the reach of traditional fixed-target and collider searches.

The relic abundance of DM is computed, where DM undergoes freeze-out in the early universe, and relic targets are drawn. The annihilation processes dominantly responsible for setting the relic abundance range from dark sector only processes in the secluded regime, to direct annihilation into SM fermions at larger portal coupling. These relic targets are compared with the current relevant constraints coming from astrophysical, laboratory, and cosmological probes.

Our primary focus is on processes giving rise to signatures at fixed-target experiments LDMX and NA64, under this reversed mass hierarchy scenario. While on-shell invisible mediator production is kinematically forbidden, we consider: off-shell dark bremsstrahlung into DM final states, dark Higgs-strahlung into dark scalar final states, invisible vector meson decays, and visible searches through decays of the $Z'$ mediator to SM. 
We compute complementary constraints from direct detection, CMB, collider searches, and BBN.
The dominant constraints come from DAMIC-M and PANDAX-4T via DM-electron scattering, collider and beam dump visible searches, and the CMB. A region of the parameter space remains unconstrained at $g_D \sim 10^{-2}$, $m_\text{DM} \gtrsim 100$ GeV.

Beyond the specific bounds of this $SU(2)_D$ model, this work highlights a broader methodological necessity in the search for light DM: the requirement to move beyond the standard on-shell mediator paradigm. By rigorously quantifying the suppression of fixed-target signatures in the inverse mass hierarchy, we demonstrate that the perceived hierarchy of sensitivity between experimental frontiers is highly model-dependent. Our findings show that direct detection is an essential primary probe that covers critical blind spots left by accelerator experiments when DM production is off-shell. This study provides a template for future dark sector explorations, underscoring that a truly robust search strategy for sub-GeV DM must treat direct detection and fixed-target probes as equal partners.

\acknowledgments
A.B. would like to thank the Chalmers University of Technology, G\"oteborg, Sweden for support during the initial stages of this work, and acknowledges support from the Department of Atomic Energy, Govt. of India. R.C. acknowledges support from an individual research grant from the Swedish Research Council (Dnr.~2022-04299). R.C. and T.G. have also been funded by the Knut and Alice Wallenberg Foundation, and performed their research within the ``Light Dark Matter'' project (Dnr. KAW 2019.0080).

\appendix

\section{Mass Matrices and Mixing Angles}
\label{section:appendix_model}

In this appendix we present the mass matrices and mxing angles in the gauge- and scalar-sector of the model. After electroweak and dark sector symmetry breaking, the neutral gauge bosons
$B_\mu$, $W^3_\mu$, and $X^0_\mu$ mix to form the physical photon $A_\mu$, the
SM-like $Z$ boson, and a new massive neutral state $Z^\prime_\mu$.
The relation between gauge and mass eigenstates is given by,
\begin{align}
    \left(\begin{array}{cc}
         B_\mu\\
         W^3_\mu \\
         X^0_\mu
    \end{array}\right) = \left(\begin{array}{ccc}
        \cos\theta_w & -\sin\theta_w\cos\rho + \tan\zeta\sin\rho& -\sin\theta_w\sin\rho - \tan\zeta\cos\rho\\
        \sin\theta_w & \cos\theta_w\cos\rho & \cos\theta_w\sin\rho\\
        0 & -\sin\rho/\cos\zeta & \cos\rho/\cos\zeta
    \end{array}\right)\left(\begin{array}{cc}
         A_\mu\\
         Z_\mu \\
         Z^\prime_\mu
    \end{array}\right).
\end{align}
The masses of the physical neutral gauge bosons are,
\begin{align}
m_Z^2
= m_W^2 A_Z
+ m_{X^0}^2 \sec^2 \zeta \sin^2 \rho\,, \quad
m_{Z'}^2
= m_W^2 A_{Z^\prime}
+ m_{X^0}^2 \sec^2 \zeta \cos^2 \rho\,,
\end{align}
where the dimensionless coefficients $A_Z$ and $A_{Z^\prime}$ are given by,
\begin{align}
\nonumber
    A_Z & = \Big[
\sec^2 \theta_w \cos^2 \rho
+ \tan \theta_w \tan \zeta
\big(
\tan \theta_w \tan \zeta \sin^2 \rho
- \sec \theta_w \sin 2\rho
\big)
\Big], \\
A_{Z^\prime} & =\Big[
\sec^2 \theta_w \sin^2 \rho
+ \tan \theta_w \tan \zeta
\big(
\tan \theta_w \tan \zeta \cos^2 \rho
+ \sec \theta_w \sin 2\rho
\big)
\Big],
\end{align}
and the parameter $\rho$ is defined as,
\begin{equation}
    \tan2\rho \equiv \frac{m_Z^2 \sin \theta_w \sin2\zeta}{m_\text{DM}^2 \cos^2\beta -m_Z^2(\cos^2\zeta-\sin^2\theta_w \sin^2 \zeta)}.
    \label{def:rho}
\end{equation}

The scalar sector of $SU(2)_D$ contains one complex doublet (4 real d.o.f.) and one real triplet (3 real d.o.f.), giving a total of 7 real scalar degrees of freedom before symmetry breaking. After spontaneous symmetry breaking, three Goldstone modes are eaten by the gauge bosons $X^a_\mu$, leaving 4 physical scalar degrees of freedom. These correspond to two neutral scalars (2 real d.o.f.) and one complex dark-charged scalar (2 real d.o.f.). After the symmetry breaking in the dark sector, the quadratic terms for the dark charged and CP-even neutral scalars are given by,
\begin{align}
    \mathcal{L}_{\rm mass} = \frac{1}{2}\left(\phi^0_D~\Sigma^0_D\right)\mathcal{M}_0^2 \left(\begin{array}{c}\phi^0_D\\\Sigma^0_D\end{array}\right) + \left(\phi^+_D~\Sigma^+_D\right)\mathcal{M}_\pm^2 \left(\begin{array}{c}\phi^-_D\\\Sigma^-_D\end{array}\right)\,,
    \label{eq:mass_terms}
\end{align}
where,
\begin{align}
    \mathcal{M}_0^2 & = \left(\begin{array}{cc} 2\lambda_{\Phi_D}v_\Phi^2 & \frac{v_\Phi}{2}(\sqrt{2}\lambda_{\Phi_D\Sigma_D}v_\Sigma-\mu_{\Phi_D\Sigma_D})\\ \frac{v_\Phi}{2}(\sqrt{2}\lambda_{\Phi_D\Sigma_D}v_\Sigma-\mu_{\Phi_D\Sigma_D}) & \lambda_{\Sigma_D} v_\Sigma^2 + \frac{\mu_{\Phi_D\Sigma_D}v_{\Phi}^2}{2\sqrt{2}v_{\Sigma}}\end{array}\right)\,,\\
    \mathcal{M}_\pm^2 & = \mu_{\Phi_D\Sigma_D}\left(\begin{array}{cc} v_\Sigma/\sqrt{2} & v_\Phi/2 \\ v_\Phi/2 & v_\Phi^2/2\sqrt{2}v_\Sigma \end{array}\right).    
\end{align}
Note that the fields in the Eq.~\eqref{eq:mass_terms} are defined in the gauge basis. The  $\chi^0$ behaves as the unphysical neutral Goldstone boson, which was eaten up by $X^0_\mu$, thus it does not appear in the mass terms. The mass matrix $\mathcal{M}^\pm$ has a zero eigenvalue, corresponding to the  Goldstone boson which acts as the longitudinal polarization of $X^\pm_\mu$. The mixing angles relating the gauge and mass eigenstates are given by the following relations,
\begin{align}
    \left(\begin{array}{cc}
         \phi^\pm_D \\
         \Sigma^\pm_D
    \end{array}\right) = \left(\begin{array}{cc}
        \cos\beta & -\sin\beta \\
        \sin\beta & \cos\beta
    \end{array}\right)\left(\begin{array}{cc}
         \hat{\phi}^\pm_D\\
         \hat{\Sigma}^\pm_D
    \end{array}\right)\, \quad \left(\begin{array}{cc}
         \phi^0_D \\
         \Sigma^0_D
    \end{array}\right) = \left(\begin{array}{cc}
        \cos\theta_t & -\sin\theta_t \\
        \sin\theta_t & \cos\theta_t
    \end{array}\right)\left(\begin{array}{cc}
         \hat{\phi}^0_D\\
         \hat{\Sigma}^0_D
    \end{array}\right),    
\end{align}
where the physical massive scalars are denoted as $\hat{\Sigma}_D^\pm$,  $\hat{\Sigma}_D^0$ and  $\hat{\phi}_D^0$, and
\begin{align}
\label{eq:theta_t}
    \tan\beta = \frac{v_\Phi}{\sqrt{2}v_\Sigma}\,, \quad \tan2\theta_t = \frac{2(\mathcal{M}^2_0)_{12}}{(\mathcal{M}^2_0)_{22}-(\mathcal{M}^2_0)_{11}}\,. 
\end{align}

\section{Decay Rates}
\label{section:appendix_CrossSections}

The decay rate of a vector meson, $V$, into vector DM through the $Z'$ is,
\begin{equation}
\begin{aligned}
    \Gamma_{V \to Z' \to X^+X^-} &=  \frac{f_V^2 g_V^2 g_D^2 \big( m_V^2 - 4m_{\text{DM}}^2 \big)^{3/2}}{192 \pi \cos^2 \zeta m_{\text{DM}}^4 
    \big( (m_{Z'}^2 - m_V^2)^2 +  \Gamma_{Z'}^2 m_{Z'}^2 \big)} \\
    & \times \Bigg[\cos^2 \rho \bigg( 4m_{\text{DM}}^2 m_V^2 ( \sin^4 \zeta - 5 \sin^2 \zeta +5 ) + m_V^4 (\sin ^2 \zeta -1)^2 + 12 m_{\text{DM}}^4 \bigg)\\
    & + 2 m_V^2 \cos \rho \sin \rho  \sin \zeta \cos \zeta  \sin \theta_w \bigg( 2 m_{\text{DM}}^2 (2 \sin^2 \zeta -5 ) + m_V^2 (\sin^2 \zeta -1)\bigg) \\
    & +  m_V^2 \sin^2\rho \sin^2 \zeta \cos^2 \zeta \sin^2 \theta_w  \bigg( 4m_{\text{DM}}^2 + m_V^2 \bigg)
    \Bigg],
\end{aligned}
\end{equation}
where $g_V = \frac{e \big( 4 \sin^2\theta \sin \rho - \sin \rho + 3 \cos \rho \tan \zeta \sin \theta_w \big)}{4 \cos \theta_w \sin \theta_w}$, where $f_V$ is the form factor for the coupling of mesons with a vector current taken from Table III of \cite{Schuster:2021mlr} which is derived from \cite{Bharucha:2015bzk}.

The invisible decay rate of the Higgs is given by,
\begin{equation}
\label{eq:InvHiggs}
\begin{aligned}
    \Gamma_{H\to Z'Z'} &= e^4 m_H \sqrt{m_H^2 - 4 m_{Z'}^2} \Big( m_H^4 - 4 m_H^2m_{Z'}^2 + 12 m_{Z'}^4 \Big) \\ 
    &\times \frac{\Big( \cos^2\theta_w \sin\rho + \sin\theta_w (\sin \rho \sin\theta_w + \cos\rho \tan\zeta) \Big)^4 v_H^2}{512 \pi \cos^4\theta_w m_{Z'}^4 \sin^2\theta_w m_H^3 },
\end{aligned}
\end{equation}
where $v_H$ is the Higgs vacuum expectation value.

\section{Feynman Rules}
\label{section:FeynmanRules}

Here we present the relevant feynman rules.

\subsection*{New Dark Gauge Bosons}
\begin{center}
\begin{minipage}{0.3\textwidth}
\centering
\includegraphics[width=0.9\linewidth]{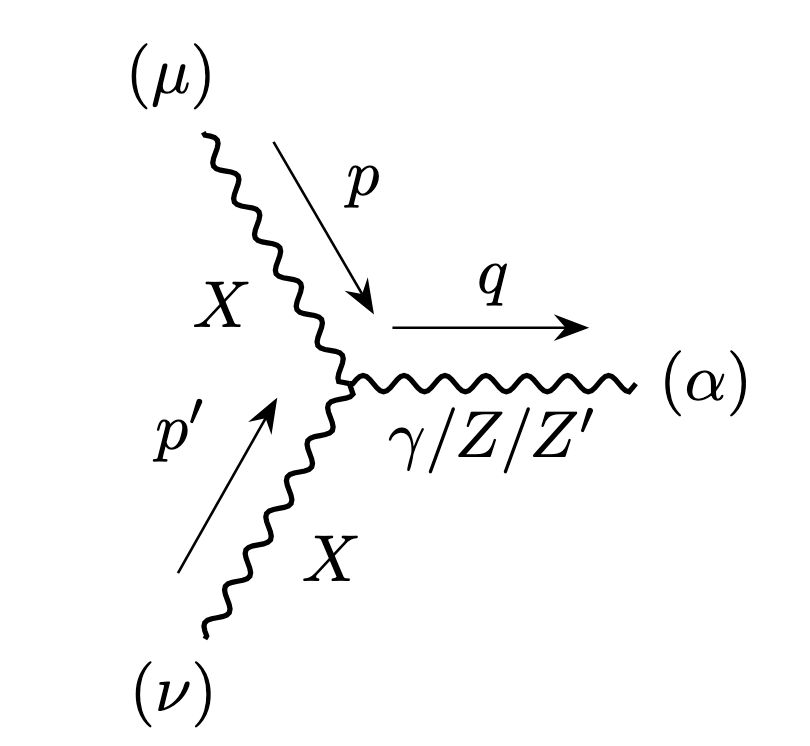}
\end{minipage}
\hfill
\begin{minipage}{0.69\textwidth}
\begin{equation}
\begin{aligned}
 = \: &i g_D \Big[ C_{1} \Big( \eta_{\mu \nu} \left( p'_{\alpha} - p_{\alpha} \right) - \eta_{\alpha \nu} \left( 2p'_{\mu} + p_{\mu} \right) + 
    \eta_{\mu \alpha} \left( p'_{\nu} + 2p_{\nu} \right) 
     \Big) \\
     &+ C_{2} \sin\zeta \Big( \eta_{\alpha\mu} q_\nu - \eta_{\alpha \nu} q_{\mu}  \Big) \Big] 
\end{aligned}
\end{equation}
\end{minipage}
\end{center}

\begin{equation}
    C_{1} = 
    \begin{cases}
        0, & \gamma \\
        \frac{\cos\rho}{\cos \zeta} , & Z' \\
        \frac{\sin\rho}{\cos \zeta} , & Z
    \end{cases}
\end{equation}

\begin{align}
    C_{2} =  \begin{cases} \cos\theta_w\,, & \gamma \\
    \left( \sin\theta_w \cos\rho + \tan\zeta \sin\rho \right)\,, & Z \\
    -\left( \sin\theta_w \sin\rho + \tan\zeta \cos\rho \right)\,, & Z^\prime \end{cases}   
\end{align}

\begin{center}
\begin{minipage}{0.3\textwidth}
\centering
\includegraphics[width=0.9\linewidth]{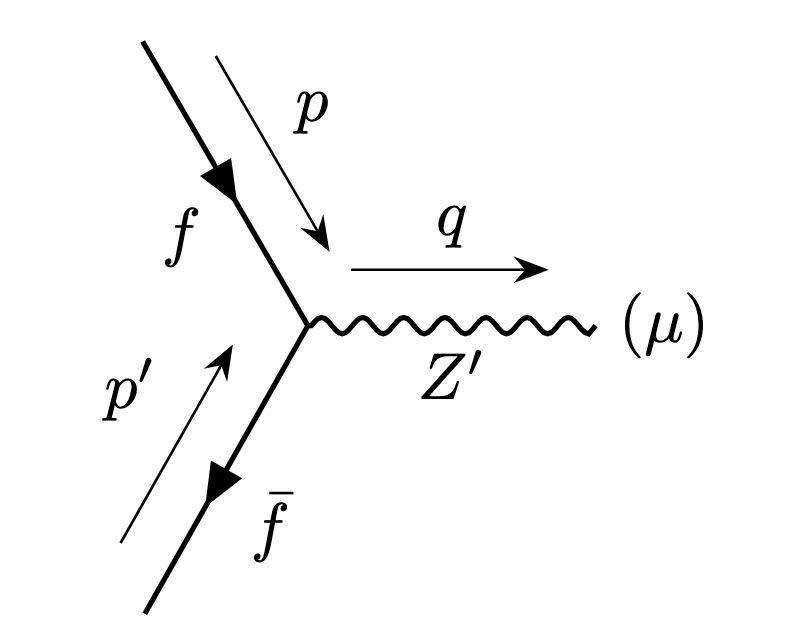}
\end{minipage}
\hfill
\begin{minipage}{0.69\textwidth}
\begin{equation}
\begin{aligned}
 = & \: \frac{ie\gamma_\mu}{4\cos \theta_w \sin \theta_w} \bigg[ \Big( \sin \rho ( 4\sin^2 \theta_w - 1) + 3 \cos \rho \tan \zeta \sin \theta_w \Big) \\ & + \gamma_5 \Big( \sin \rho+ \cos \rho \tan \zeta \sin \theta_w \Big) \bigg]
 \label{eq:FRZ'ff} 
 \end{aligned}
\end{equation}
\end{minipage}
\end{center}

\subsection*{New Scalar Interactions}
\begin{center}
\begin{minipage}{0.3\textwidth}
\centering
\includegraphics[width=0.9\linewidth]{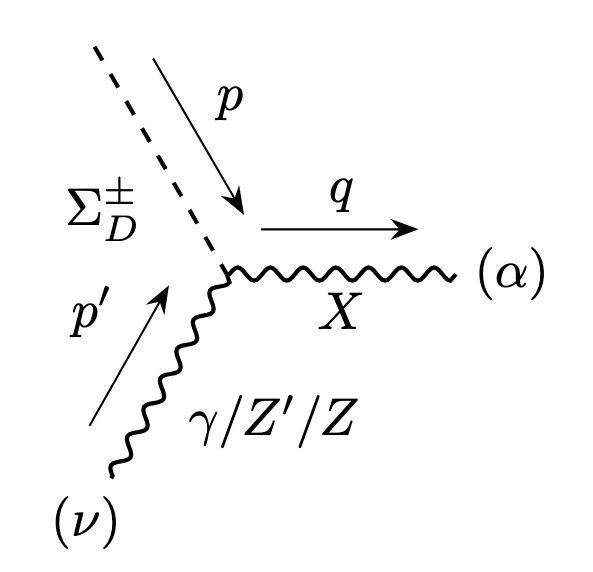}
\end{minipage}
\hfill
\begin{minipage}{0.69\textwidth}
\begin{equation}
 = \: \frac{i g_D \cos \beta \sin \zeta}{ m_{\text{DM}} \sin \beta} \Big[ C_{1,S} \eta_{\nu \alpha} - C_{2,S} q_\nu p'_\alpha \Big] 
\end{equation}
\end{minipage}
\end{center}
\begin{equation}
    C_{1,S} = 
    \begin{cases} 
    \cos \theta_w (p' \cdot q), &  \gamma \, \\
    \frac{4 m_\text{DM}^2 \sin \rho \sin^2 \beta}{\cos \zeta \sin \zeta} +(p' \cdot q) \left(  \tan\zeta \sin\rho - \sin\theta_w \cos\rho \right)\,, & Z\, \\
    -\frac{4 m_\text{DM}^2 \sin \rho \sin^2 \beta}{\cos \zeta \sin \zeta} - (p' \cdot q) \left(  \tan\zeta \cos \rho + \sin\theta_w \sin\rho \right)\,, & Z'\, 
    \end{cases}
\end{equation}

\begin{equation}
    C_{2,S} = 
    \begin{cases} 
    \cos \theta_w, & \gamma \, \\
   \left( \cos\rho \sin \theta_w - \sin\rho \tan \zeta \right)\,, & Z\, \\
    \left( \sin\rho \sin \theta_w + \cos\rho \tan \zeta \right) \,, &Z'\, 
    \end{cases}
\end{equation}

\begin{center}
\begin{minipage}{0.3\textwidth}
\centering
\includegraphics[width=0.9\linewidth]{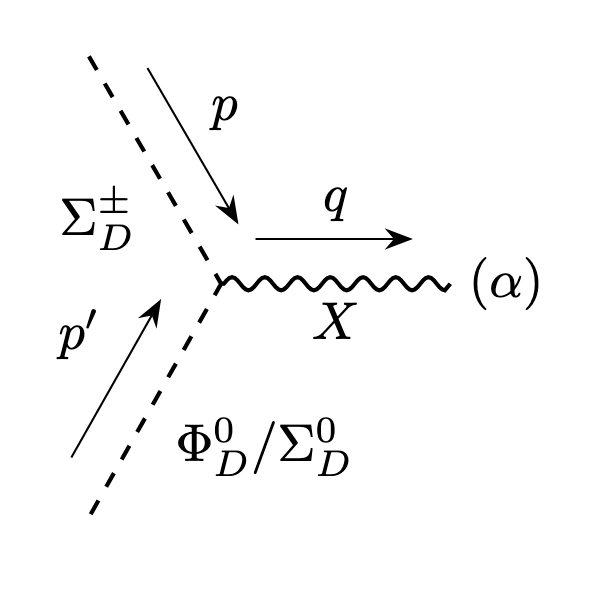}
\end{minipage}
\hfill
\begin{minipage}{0.69\textwidth}
\begin{equation}
\begin{aligned}
 = i g_D \Big( p'_\alpha - p_\alpha \Big) \times
 \begin{cases}
     \big( \cos \theta_t \sin \beta + 2\sin\theta_t \cos \beta \big), & \Phi_D^0 \\
     \big( \sin \theta_t \sin \beta - 2\cos\theta_t \cos \beta \big), & \Sigma_D^0
 \end{cases}
\end{aligned}
\end{equation}
\end{minipage}
\end{center}

\begin{center}
\begin{minipage}{0.3\textwidth}
\centering
\includegraphics[width=0.9\linewidth]{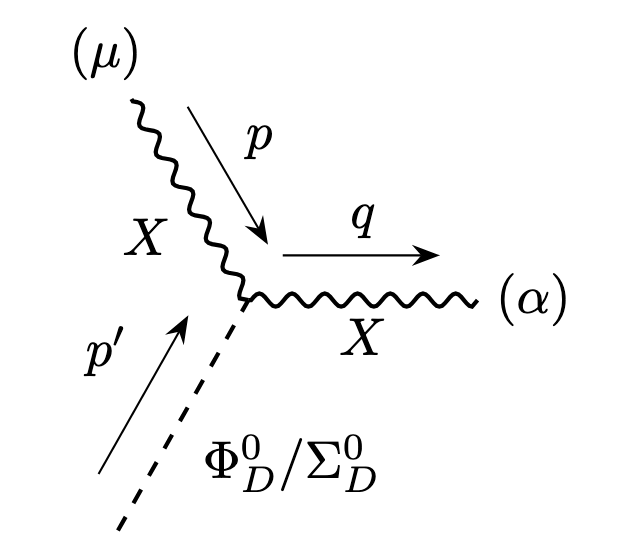}
\end{minipage}
\hfill
\begin{minipage}{0.69\textwidth}
\begin{equation}
\begin{aligned}
= i g_D m_\text{DM} \cos \beta \eta_{\mu \alpha} \times 
 \begin{cases}
     \big( \cos \theta_t - 2\sin\theta_t \big), & \Phi_D^0 \\
     \big( \sin \theta_t \sin \beta + 2\cos\theta_t \big), & \Sigma_D^0
 \end{cases}
\end{aligned}
\end{equation}
\end{minipage}
\end{center}

\begin{center}
\begin{minipage}{0.3\textwidth}
\centering
\includegraphics[width=0.9\linewidth]{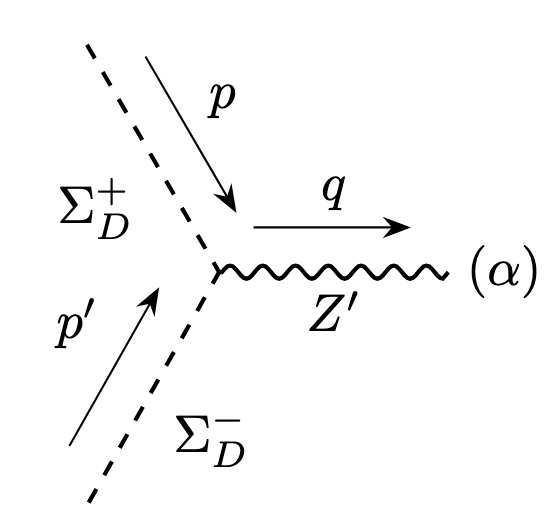}
\end{minipage}
\hfill
\begin{minipage}{0.69\textwidth}
\begin{equation}
\begin{aligned}
 = \: i g_D \frac{\cos \rho}{\cos \zeta} \big( \cos^2\beta + \sin^2\beta/2 \big) \big(p_\alpha-p'_\alpha\big)
\end{aligned}
\end{equation}
\end{minipage}
\end{center}

\begin{center}
\begin{minipage}{0.3\textwidth}
\centering
\includegraphics[width=0.9\linewidth]{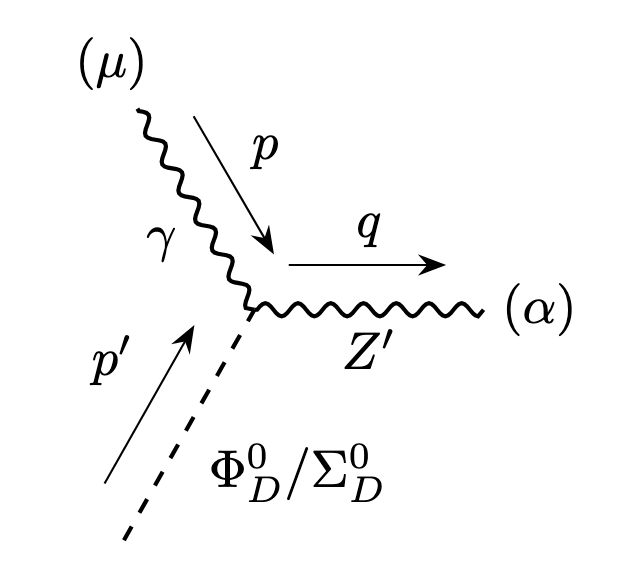}
\end{minipage}
\hfill
\begin{minipage}{0.69\textwidth}
\begin{equation}
\begin{aligned}
 = ig_D\frac{\cos \rho \cos \theta_w \sin \zeta}{m_\text{DM} \cos\zeta \sin \beta} \Big( p_\alpha q_\mu - \eta_{\alpha \mu} (p \cdot q) \Big) \times
 \begin{cases}
     \sin \theta_t, & \Phi_D^0 \\
     -\cos \theta_t, & \Sigma_D^0
 \end{cases}
\end{aligned}
\end{equation}
\end{minipage}
\end{center}

\section{Relic Density Processes}
\label{section:relic_density_processes}

As described in \secref{section:RelicDensity}, the freeze-out relic density is dominantly set by the two main DM annihilation processes in the early universe: DM annihilation into SM fermions corresponding to the \textit{direct annihilation} regime, and into $Z'$s corresponding to the \textit{secluded annihilation} regime.
\begin{figure}[h]
    \centering
    \includegraphics[width=0.49\linewidth]{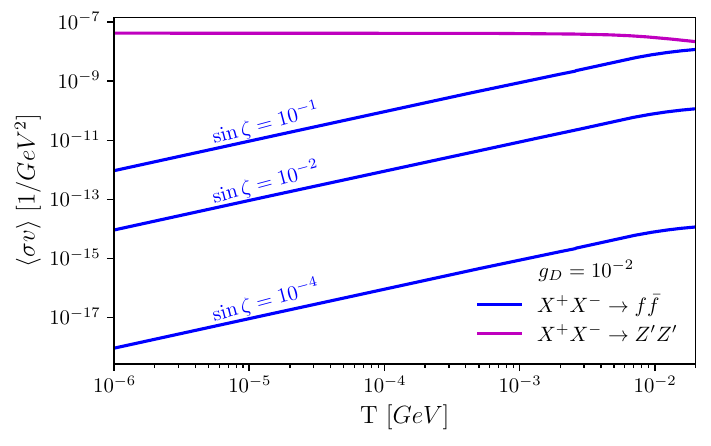}
    \includegraphics[width=0.49\linewidth]{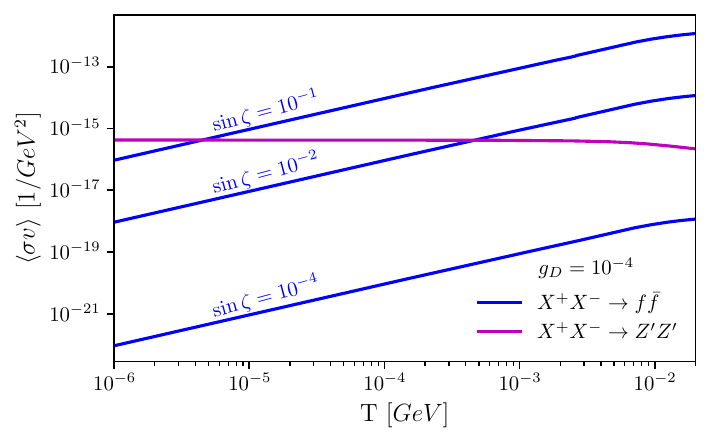}
    \caption{Thermally averaged annihilation cross sections $\langle \sigma v \rangle$ as a 
function of temperature $T$ for the two dominant DM freeze-out processes, computed with 
$m_\text{DM} = 0.1$ GeV. The direct annihilation process $X^+X^- \to f\bar{f}$ (blue) 
is shown for three values of the kinetic mixing angle, $\sin\zeta \in \{10^{-1}, 10^{-2}, 
10^{-4}\}$, while the secluded annihilation process $X^+X^- \to Z'Z'$ (magenta) is 
independent of $\sin\zeta$. The left panel corresponds to $g_D = 10^{-2}$ and the right 
to $g_D = 10^{-4}$. Freeze-out occurs at $T_\text{FO} \sim m_\text{DM}/10 = 0.01$ GeV 
(right edge of the plots), where the process with the largest $\langle \sigma v \rangle$ 
dominates the relic density.}
    \label{fig:relic_processes}
\end{figure}
 The thermally averaged cross sections, $\langle \sigma v \rangle$, are plotted as a function of temperature for each process in \figref{fig:relic_processes} for two values of $g_D$.
The left corresponds to $g_D = 10^{-2}$ and right to $g_D=10^{-4}$, with $m_\text{DM} = 0.1$ GeV for both. The direct annihilation process into SM fermions is plotted for various values of $\sin \zeta$ in blue, while the secluded process remains independent of $\sin \zeta$ and is plotted in magenta. In the left panel, the secluded process dominates for small $\sin\zeta$, while direct annihilation takes over for $\sin\zeta \gtrsim 10^{-2}$. In the right panel, the $g_D^4$ suppression of the secluded channel renders it subdominant 
except for $\sin\zeta \lesssim 10^{-2}$, illustrating the sensitivity of the regime 
transition to $g_D$.

Freeze-out occurs at $T \sim m_\text{DM}/10 = 0.01$ GeV, thus the process with the largest $\langle \sigma v \rangle$ at this temperature is the process dominantly responsible for setting the relic density.
For a given $g_D$ and $m_\text{DM}$, above a certain value of $\sin \zeta$ the direct annihilation process becomes the dominant process during freeze-out. This corresponds to the change in slope of the relic target in \figref{fig:Sensitivity}. 
In the secluded regime, $\Omega h^2 \propto 1/\langle\sigma v\rangle_\text{secluded} \propto g_D^{-4}$, while in the direct regime $\Omega h^2 \propto g_D^{-2} \sin^{-2}\zeta$. The crossover between regimes therefore occurs at a critical mixing angle $\sin\zeta^* \propto g_D$.
The cross section is inversely proportional to the DM mass, therefore increasing $g_D$ causes the relic targets to shift to larger $m_\text{DM}$.

\bibliographystyle{JHEP}
\bibliography{references}
\end{document}